\documentclass{aa}
\usepackage{graphicx}
\usepackage{txfonts}
\usepackage{natbib}
\bibpunct{(}{)}{;}{a}{}{,}

\newcommand{\be}{\begin{enumerate}}
\newcommand{\ee}{\end{enumerate}}

\def\mathstacksym#1#2#3#4#5{\def#1{\mathrel{\hbox to 0pt{\lower#5\hbox{#3}\hss} \raise #4\hbox{#2}}}}
\mathstacksym\gta{$>$}{$\sim$}{1.5pt}{3.5pt} 
\mathstacksym\lta{$<$}{$\sim$}{1.5pt}{3.5pt} 
\begin{document}
                                %
\title{Exploring the Lower Mass Function in the young open Cluster
  IC\,4665\thanks{Based on observations obtained at the Canada-France-Hawaii Telescope (CFHT) which is operated by the National Research Council of Canada, the Institut National des Science de l'Univers of the Centre National de la Recherche Scientifique of France,
and the University of Hawaii.}}
\titlerunning{Young open cluster IC\,4665}
\author{        W.J. de Wit\inst{1} \and
  J. Bouvier\inst{1} \and
  F. Palla\inst{2} \and
  J.-C. Cuillandre\inst{3}\and
  D.J. James\inst{4}\and
  T.R. Kendall\inst{5} \and
  N. Lodieu\inst{6,8}\and
  M.J. McCaughrean\inst{7,8}
  E. Moraux\inst{9}\and
  S. Randich\inst{2}\and
  L. Testi\inst{2}
} 
\offprints{W.J. de Wit, \email{dewit@obs.ujf-grenoble.fr}}
\institute{Laboratoire d'Astrophysique, Observatoire de Grenoble, BP 53, 38041 Grenoble, C\'{e}dex 9, France\and 
          INAF, Osservatorio Astrofisico di Arcetri, Largo E. Fermi 5, 50125 Florence, Italy \and                        
          Canada-France-Hawaii Telescope Corp., Kamuela, HI 96743, USA\and
          Physics and Astronomy Department, Vanderbilt University, 1807 Station B, Nashville, TN 37235, USA\and
          Centre for Astrophysics Research, Science \& Technology Research
          Institute, Department of Physics, Astronomy \& Mathematics, University of Hertfordshire, College Lane, Hatfield AL10 9AB, UK\and
          University of Leicester, Department of Physics \& Astronomy, University Road, Leicester LE1 7RH, UK\and
          University of Exeter, School of Physics, Stocker Road, Exeter EX4 4QL, UK\and
          Astrophysikalisches Institut Potsdam, An der Sternwarte 16, 1448 Potsdam, Germany\and
          Institute of Astronomy, Cambridge, CB3 0HA, UK
}          
        
\date{Received date; accepted date}
\abstract{We present a study of the young (30-100\,Myr) open cluster IC\,4665
  with the aim to determine the shape of the mass function well into the brown
  dwarf regime. We photometrically select 691 low-mass stellar and 94 brown
  dwarf candidate members over an area of 3.82 square degrees centred on the
  cluster.  $K$-band follow-up photometry and Two-Micron All-Sky Survey data
  allow a first filtering of contaminant objects from our catalogues. A second
  filtering is performed for the brightest stars using proper motion data
  provided by the Tycho-2 and UCAC2 public catalogues. Contamination by the
  field population for the lowest mass objects is estimated using same latitude
  control fields. We fit observed surface densities of various cluster
  populations with King profiles and find a consistent tidal radius of
  $1.0^{\circ}$. The presence of possible mass segregation is discussed. In most
  respects investigated, IC\,4665 is similar to other young open clusters at
  this age: (1) a power law fit to the mass function between 1 and
  $0.04\,M_{\odot}$ results in best fit for a slope of $-0.6$; (2) a cusp
  in the mass function is noticed at about the substellar boundary with respect
  to the power law description, the interpretation of which is discussed; (3) a
  fraction between 10-19\% for BDs with $M \gta 0.03\,M_{\odot}$ to total members;
  (4) a best-fit lognormal function to the full mass distribution shows an
  average member mass of $0.32\,M_{\odot}$, if IC\,4665 has an age of 50\,Myr.
   }
\maketitle
\section{Introduction}
Formation theories envisage brown dwarfs (BDs) evolving from the
gravitational collapse of the smallest unities in the opacity limited
fragmentation of giant molecular clouds like stars,
e.g. \citet{1987ARA&A..25...23S}. Otherwise they may form from the condensation
of solids within circumstellar (or circumbinary) disks like planets
\citep{1998Sci...281.2025L, 2001MNRAS.325..221P,
2004AJ....127..455J}. Innovative ideas let BDs undergo a star-like gravitational
collapse initially, but they terminate the subsequent mass accretion on the
formed ``protostar''. This could be due to the removal of the protostar from the
material reservoir by a dynamical ejection through a dynamical interaction
with one of the its siblings \citep{2001AJ....122..432R}. Conversely the
growth of the stellar seed can be halted by the removal of the material
reservoir itself either due to the intense radiation pressure from nearby young
high-mass stars \citep{2001MNRAS.322..231K,2004A&A...427..299W} or due to the
actual collision of protostars, by which the impact causes a large portion of
the envelope to be removed.  As such these various ideas supply testable observational
predictions like ``BD halos'' of young clusters, a relative proximity to massive
stars or a regular mixture with low mass stars. They will lead to an
increased understanding of BDs and the star formation (SF) process in general. 

Deriving the underlying initial mass function (IMF) is generally considered as
an outstanding tool for revealing the physical mechanisms at play.  It is only for a
handful regions that one has been able to derive the mass distribution crossing
the sub-stellar boundary at $0.072\,M_{\odot}$, \citep[see][ and references
therein]{2003IAUS..211..147B}.  The mass function of young open clusters with an
age of $\rm \sim100\,Myr$ seems best described by a lognormal law over the full
mass range in a dN/dlogM presentation, regardless of their exact age, metallicity or richness. On the contrary,
young regions with active SF ($\rm \lta 3\,Myr$) may indicate a mutually discordant
number of BDs per star \citep{2003ApJ...593.1093L}.  Additionally the binary
frequency among field BDs indicates that the formation of BD binaries cannot be a mere
extension of the formation of binary low mass stars
\citep{2003ApJ...587..407C}. As argued by \cite{2003MNRAS.346..369K}, it would
indicate that BDs do not follow the same general rules for formation as stars
do. On the other hand, BDs are known to be
accompanied by CS disks \citep{2002A&A...393..597N,2004A&A...424..603N}. Mohanty et al. (2005)\nocite{2005ApJ...626..498M} specifically demonstrate
that young BDs do experience a T\,Tauri-like phase. These findings combined with
the recent detection of a wide ($\rm >200\,AU$) BD binary (Luhman
2004)\nocite{2004ApJ...614..398L}, and the similar velocity distribution for stars and BDs in
synthetic small-N clusters (Delgado-Donate et al. 2004\nocite{2004MNRAS.347..759D}) all argue in
favor of BD formation being very much similar to low-mass star formation. It is
clear that the dominant process that determines the final mass
of a low mass star or a BD is not settled upon.

\begin{figure}
  \includegraphics[height=8cm,width=8cm]{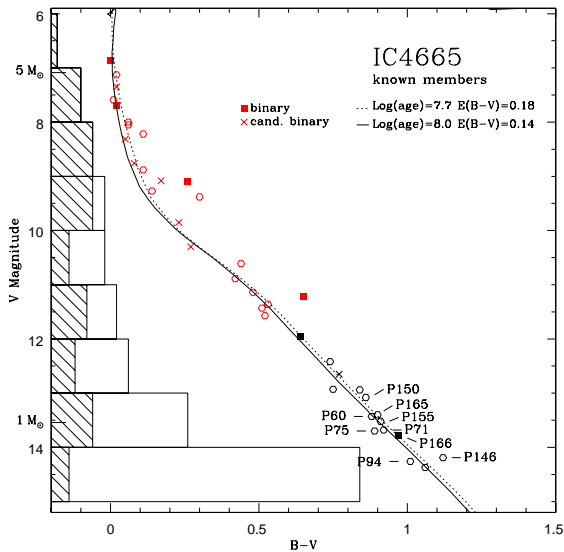}
  \caption[]{Known member stars of IC\,4665.  Two
    isochrones for 50\,Myr and 100\,Myr with indicated reddening (for $ R_{\rm
    v}=3.1$) from the Padua group (Girardi et
    al. 2000\nocite{2000A&AS..141..371G}). Binary and candidate binary systems
    have specific symbols. Some members are named (see also Fig.\,\ref{binar}). The
    plain histogram represents relative numbers in 
    each magnitude bin of candidate members (Prosser 1993); the hatched part the
    confirmed members. The masses on the y-axis correspond to the 50\,Myr
    isochrone.}
  \label{prevmem}
\end{figure}

Young open clusters that occupy the age interval of 10 - 50\,Myr (hereafter
called pre-main sequence clusters or PMS clusters) can provide the clues to the
issues listed above. There are a number of reasons why this age interval is
critical in deriving the sub-stellar IMF.  Unlike SF regions, PMS clusters
provide the sampling of a complete stellar population immediately after the end
of the active star forming phase, thus revealing the full stellar
product. Secondly BDs in PMS clusters are still bright and easily detected and
in addition they do not suffer from prodigious and inhomogeneous extinction as
is generally the case for SF regions. From a methodological point of view, the
IMF is obviously not what is observed, but is what is obtained after
transformation of the luminosity function. The derived IMF thus depends on the
theoretical predictions for the dependence of magnitude on mass and age (and
metallicity).  \citet{2002A&A...382..563B} emphasized that no current BD model
can be relied upon for ages less than 1\,Myr, due to uncertainties in the
treatment of convection, mass accretion rate, and atmospheric parameters ($\rm
H_{2}$ abundance). It is therefore that mass functions derived for somewhat
older regions, like PMS clusters that will play a more accurate role than the MF
derived for young SF regions. Clusters older than $\rm \sim 150\,Myr$ start to
be significantly affected by dynamical evolution, leading to serious problems in
deriving the present-day MF. An additional complicating factor for these somewhat
older clusters is the intrinsically fainter BDs. Clusters in the PMS
age interval are therefore well-suited to explore in a relatively unbiased way 
the MF and thus to constrain BD formation theories, provided that a large
enough area, probing the full range of cluster radii from centre to edge is covered. Unfortunately, only a few PMS
clusters do actually exist in the solar neighbourhood ($\rm < 500\,pc$).

IC\,4665 is one of the few young open (and possibly PMS) clusters within the solar
neighbourhood, at a distance of $\rm \sim 350\,pc$ ($\rm D_{Hipp}=385\pm40\,pc$,
Hoogerwerf et al. 2001\nocite{2001A&A...365...49H}).   Previous work was undertaken primarily by
Prosser and collaborators, scrutinizing candidate members on spectral type,
Li-abundance, proper motion, radial velocity, H$\alpha$-emission and X-rays with
ROSAT (Prosser 1993\nocite{1993AJ....105.1441P}; Prosser \& Giampapa
1994\nocite{1994AJ....108..964P}; Mart\'{\i}n \& Montes
1997\nocite{1997A&A...318..805M}; Giampapa, Prosser \& Fleming
1998\nocite{1998ApJ...501..624G}).  An age estimate of IC\,4665 by Mermilliod
(1981\nocite{1981A&A....97..235M}) indicated an isochronal age of 30-40\,Myr
based on an analysis of the upper main-sequence stars.  However Prosser (1993)
argued that due to membership uncertainties, photometric errors and a colour
effect of high rotational velocities, the upper main sequence stars are
consistent with the age of \object{$\alpha$ Per} and may even indicate an age
similar to the Pleiades.
The age of IC\,4665 is therefore not firmly established, but should lie
somewhere between 30\,Myr and 100\,Myr. By definition IC\,4665 is possibly not a
PMS cluster. The cluster sequence is well fitted
both by a 50 Myr and 100 Myr isochrone (see Fig.\,\ref{prevmem}). 
Although located out of the Galactic plane at a latitude of
$b=+17^{o}$ a significant background stellar population does exist in that
specific direction. We refer to Prosser (1993) for an overview of the work by
various authors prior to 1990. In Fig.\,\ref{prevmem}, we summarize current
membership of IC\,4665 prior to the work presented here in the form of a
colour-magnitude diagram. The listed characteristics of IC\,4665 combined with
its rather well-known intermediate mass and solar-mass members render the
cluster an excellent target for probing the IMF into the regime of very low mass
stars and brown dwarfs.

This paper is organized as follows. We give a brief overview of the
observations, data reduction and calibration in
Sect.\,\ref{obsdat}. Sect.\,\ref{newm} describes the procedure for selecting new
candidate members reaching well below the hydrogen burning limit. The fractional
contamination by non-member objects is estimated in Sect.\,\ref{cont} using
control fields, $K$-band photometry, and proper motion. We derive and discuss
the radial distribution, the mass function and the total mass of the cluster in
Sect.\,\ref{disc}. A summary of the work presented is given in
Sect.\,\ref{summary}.

\begin{figure}
  \includegraphics[height=8cm,width=8cm]{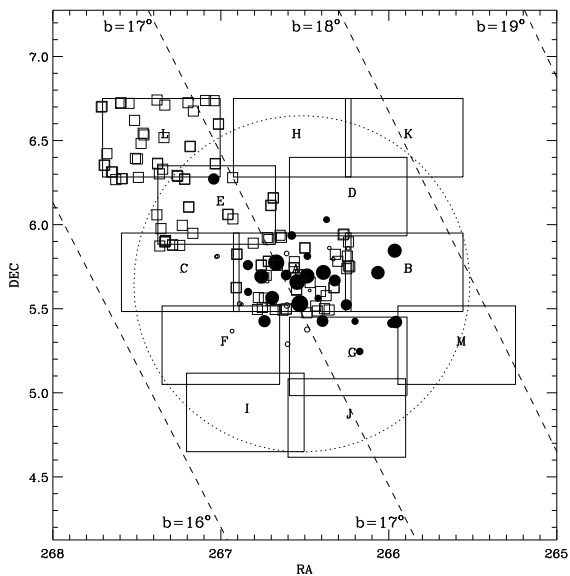}
  \caption[]{Thirteen CFHT12k fields (large boxes) covering 3.82 square degrees of IC\,4665 were
    observed. Note the overlaps between the various fields. The small boxes correspond to the fields observed
    with CFHT-IR. IC\,4665 members are represented by the large filled dots,
    whose sizes are in linear relation to the $V$-band magnitude. Finally, the large dashed circle
    corresponds to a tidal radius for the best fit King
    profile (see Sect.\,\ref{disc}).}
  \label{spat}
\end{figure}
\begin{table}[t]
  {
    \begin{center}
      \caption[]{Central coordinates of IC4665 pointings (fields) with CFHT using CFH12K
        camera.}
      \begin{tabular}{cccc}
        \hline
        \hline
        Field & RA (2000) & Dec (2000)& Date \\
              & (h,m,s)   & ($\degr$,$\arcmin$,$\arcsec$) & \\
        \hline
        A &17:46:18 & +05:43:00  & 19-5-2002\\
        B &17:43:38 & +05:43:00  & 20-5-2002\\
        C &17:48:58 & +05:43:00  & 20-5-2002\\
        D &17:44:58 & +06:10:00  & 20-5-2002\\
        E &17:48:06 & +06:07:00  & 20-5-2002\\
        F &17:48:00 & +05:17:00  & 09-6-2002\\
        G &17:44:58 & +05:13:00  & 09-6-2002\\
        H &17:46:18 & +06:31:00  & 10-6-2002\\
        I &17:47:25 & +04:53:00  & 10-6-2002\\
        J &17:45:00 & +04:51:00  & 09-6-2002\\
        K &17:43:38 & +06:31:00  & 09-6-2002\\
        L &17:49:25 & +06:31:00  & 10-6-2002\\
        M &17:42:23 & +05:17:00  & 09-6-2002\\
\hline
       C1 &17:51:03 & +08:05:54  & 12-6-2002\\
       C2 &17:40:53 & +02:50:14  & 12-6-2002\\
        \hline
      \end{tabular}
      \label{obslog}
    \end{center}
  }
\end{table}

\section{Observations and data reduction}
\label{obsdat}
\subsection{Optical photometry}
As part of a CFHT large program surveying star forming regions and young open
clusters, IC\,4665 was observed in the light of the I (Mould) and $z$
filters. The telescope at that time was equipped with the CFH12K mosaic camera,
the predecessor of the wide-field imager MegaCam. We refer to
\citet{2003A&A...400..891M} for more details regarding observations, data
reduction and astrometric calibration of the CFHT 12k camera mosaic. In what
follows we describe some specific calibration steps applied to the IC\,4665 set
of observations.

The CFH12K camera consists of 12 individual $2048\times4096$ CCDs each with
angular scale of 0.206$^{\prime\prime}$\,pix$^{-1}$ arranged in a $2\times6$
mosaic \citep{2001sf2a.conf..605C}. In total 13 CFH12K pointings (named field A to M)
centred on IC\,4655 were observed for a total of 3.82 square degrees. Two control
fields, located at a three degrees from IC\,4665 at the same Galactic latitude were observed in order to
estimate the foreground and background contamination of field stars. The central
coordinate and the observation dates of each field are given in overview in
Table\,\ref{obslog}. The coordinates of the fields were chosen such that each
mosaic would overlap another. The overlaps are used to determine the external
error on the photometry and allows the creation of an internal photometric
system by calibrating the fields to one ``master'' field. The projection on the
sky of the covered area relative to the brightest members of IC\,4665 is shown
in Fig.\,\ref{spat}.  Each of the 13 fields was observed for three different
exposure times: 2s, 30s and 300s or 360s in I and $z$ band. It allows the
sampling of member stars for a range in masses between approximately 1 and
0.01\,$\rm M_{\odot}$.

\begin{figure}
  \includegraphics[height=8cm,width=4cm,angle=90]{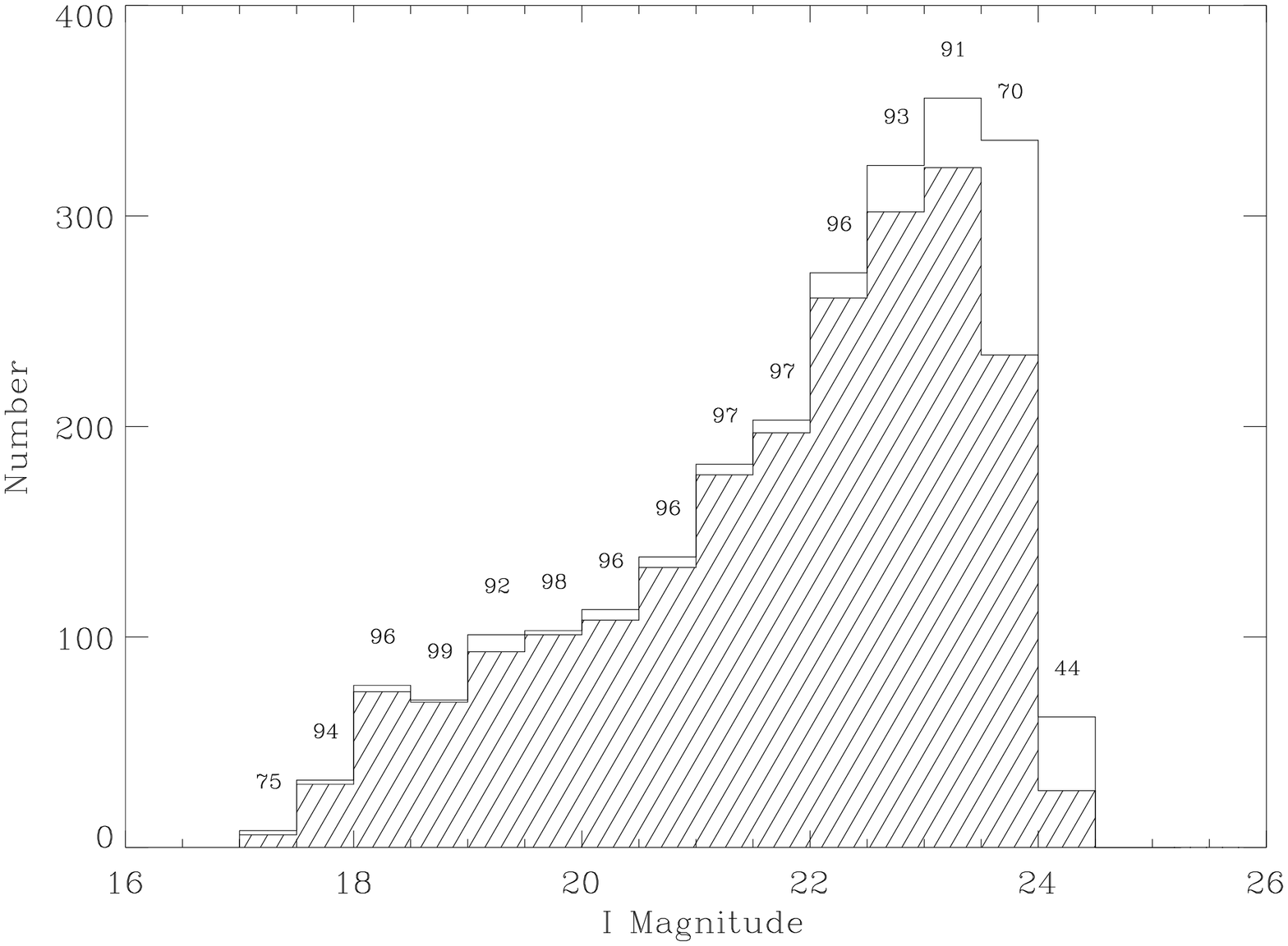}
  \includegraphics[height=8cm,width=4cm,angle=90]{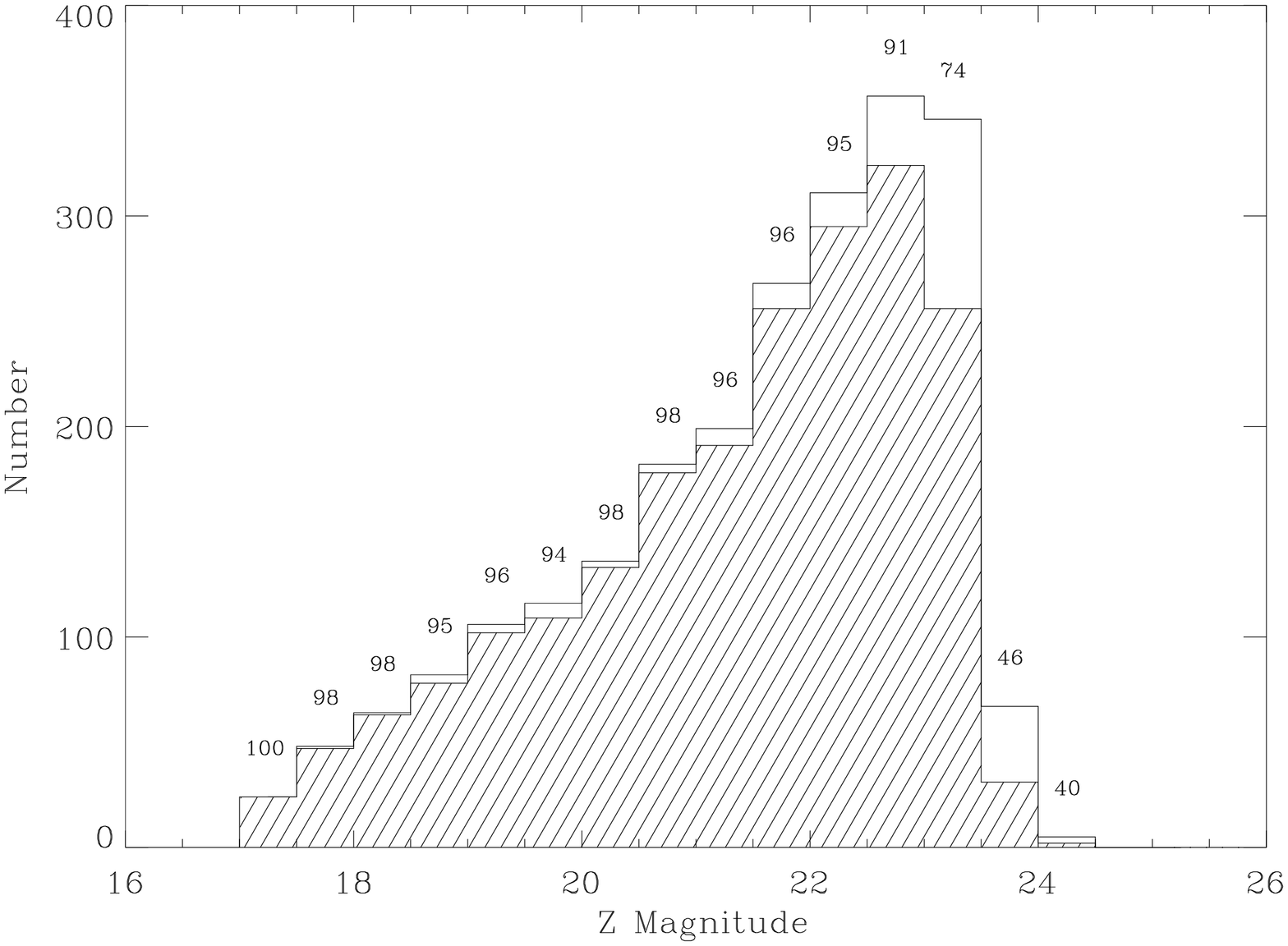}
  \caption[]{Histogram of objects in field 'D' located in the overlap with field
  'K'. The hashed histogram presents objects detected in both fields, and
  measures  the completeness at each magnitude bin.}
  \label{compl}
\end{figure}

Objects were extracted from the CCD frames using the Sextractor software package
(Bertin \& Arnouts 1996\nocite{1996A&AS..117..393B}). To increase the accuracy
on the photometric measurements, we used sextractor in conjunction with a point
spread function (PSF) source extraction procedure. The PSF is determined for
each individual CCD of the CFH12K mosaic for each field by the most well-behaved
stars using the PSFex software (Bertin, priv. comm.; see also Kalirai et al. 2001\nocite{2001AJ....122..257K}). The procedure
obviously requires a constant PSF profile across a CCD. A changing PSF profile
however was observed on the $I$-band images for each CCD of the CFH12K mosaic and
for each of the three exposures. The PSF profile was constant as function of
X-coordinate, but changed as function of Y-coordinate: at Y-coordinates less
than $\sim 500$ the PSF profile turns into an asymmetric, ``teardrop''-like shape. This instrumental
effect becomes especially apparent when comparing magnitudes obtained with
aperture photometry and PSFex photometry. Such a comparison reveals an
underestimation of the $I$-band magnitude and reaches a maximum $0.2^{m}$ for PSF
extracted stars at Y-coordinates less 50. This is due to the loss of flux by the
difference in shape of the actual stellar PSF and the applied
model-PSF. Surprisingly, images obtained in the $z$-band do not suffer from
variable PSF profile, hinting at an $I$-band filter problem of the CFH telescope.
As far as we know, this is the first time this problem is reported for CFH12K
$I$-band images. The immediate consequence of the variable PSF is that objects at
low Y-coordinates on the CCD appear redder than they actually are. The PSF
effect introduces therefore a higher fractional contamination of non-members to
our photometrically selected candidate member catalogues, but does not lead to
detection failure of IC\,4665 members. On the other hand, magnitudes determined
from PSF fitting are more accurate than aperture photometry. We therefore chose
to continue using the PSF magnitudes, however flagging candidate member stars
when found at low Y-coordinates. The completeness of the survey is exemplified
in Fig.\,\ref{compl}, after applying nominal zeropoints as determined by
the CFHT's Elixir pipe-line (Magnier \& Cuillandre)
2004\nocite{2004PASP..116..449M}. These limits have been determined from the
stars detected in the overlapping region of the 300s exposure between field 'D' and 'K'. The figure shows that I and z band
are 90\% complete down to a magnitude of $\sim 23^{m}$; the $I$-band is about
$0.5^{m}$ deeper than the $z$-band.

\begin{figure}
  \includegraphics[height=8cm,width=8cm]{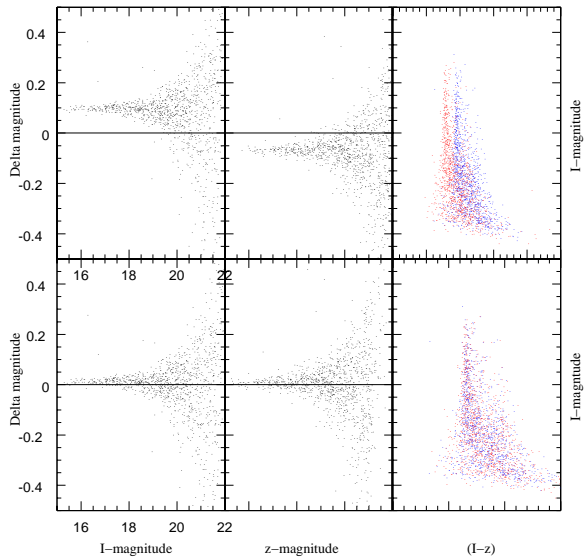}
  \caption[]{Exemplifying the systematic shift between observations of the same
    part of the sky but with different CCDs of the CFH12K mosaic. The left
    ($I$-band) and middle ($z$-band) columns show CCD06 of field\,I minus CCD10 of field\,J in a
    30s exposure, before (upper row) and after (lower row) applying a
    zeropoint shift. The effect in a CMD (right column) is substantial.}
  \label{fshift}
\end{figure}

\begin{figure}
  \includegraphics[height=8cm,width=4cm,angle=90]{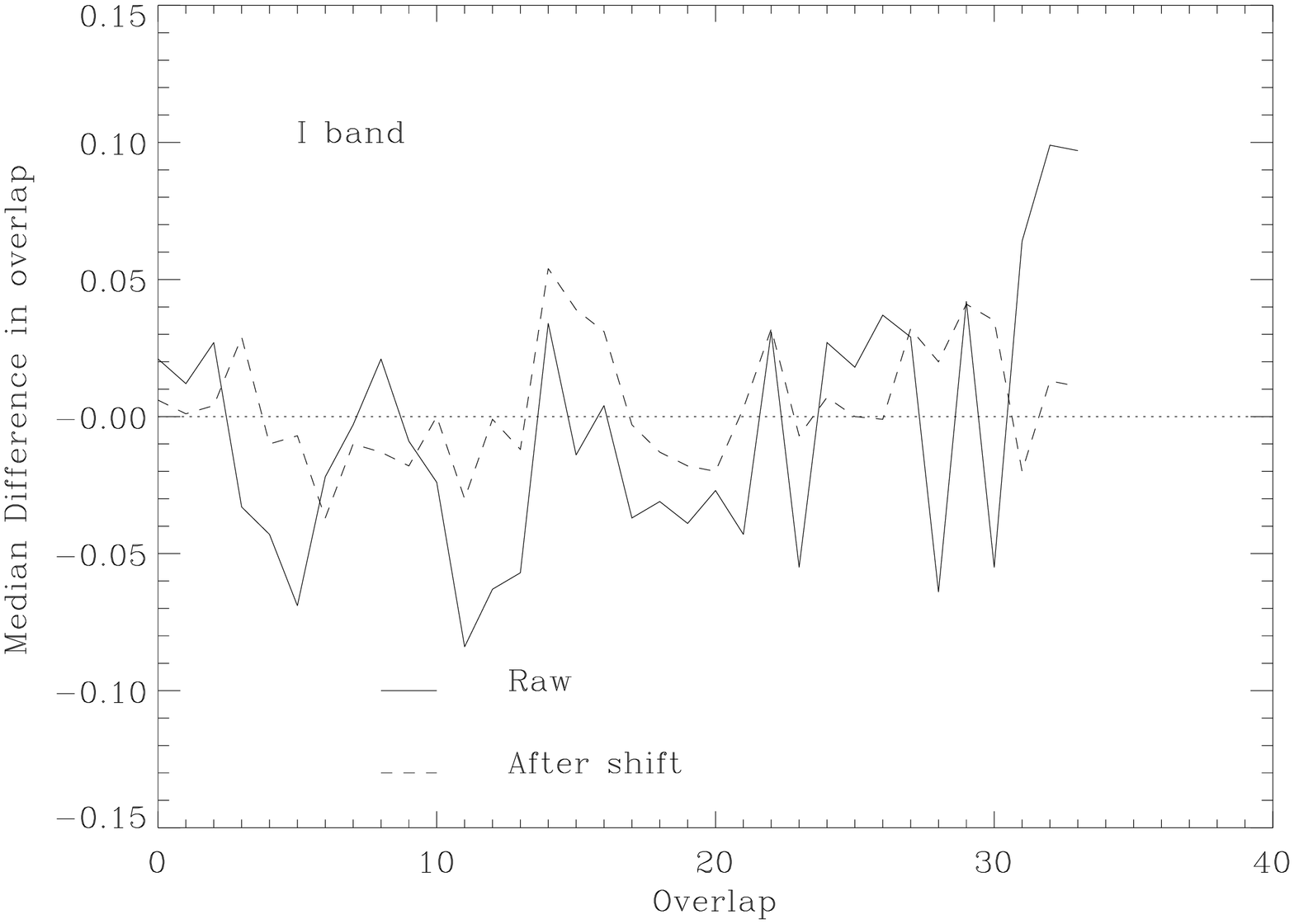}
  \includegraphics[height=8cm,width=4cm,angle=90]{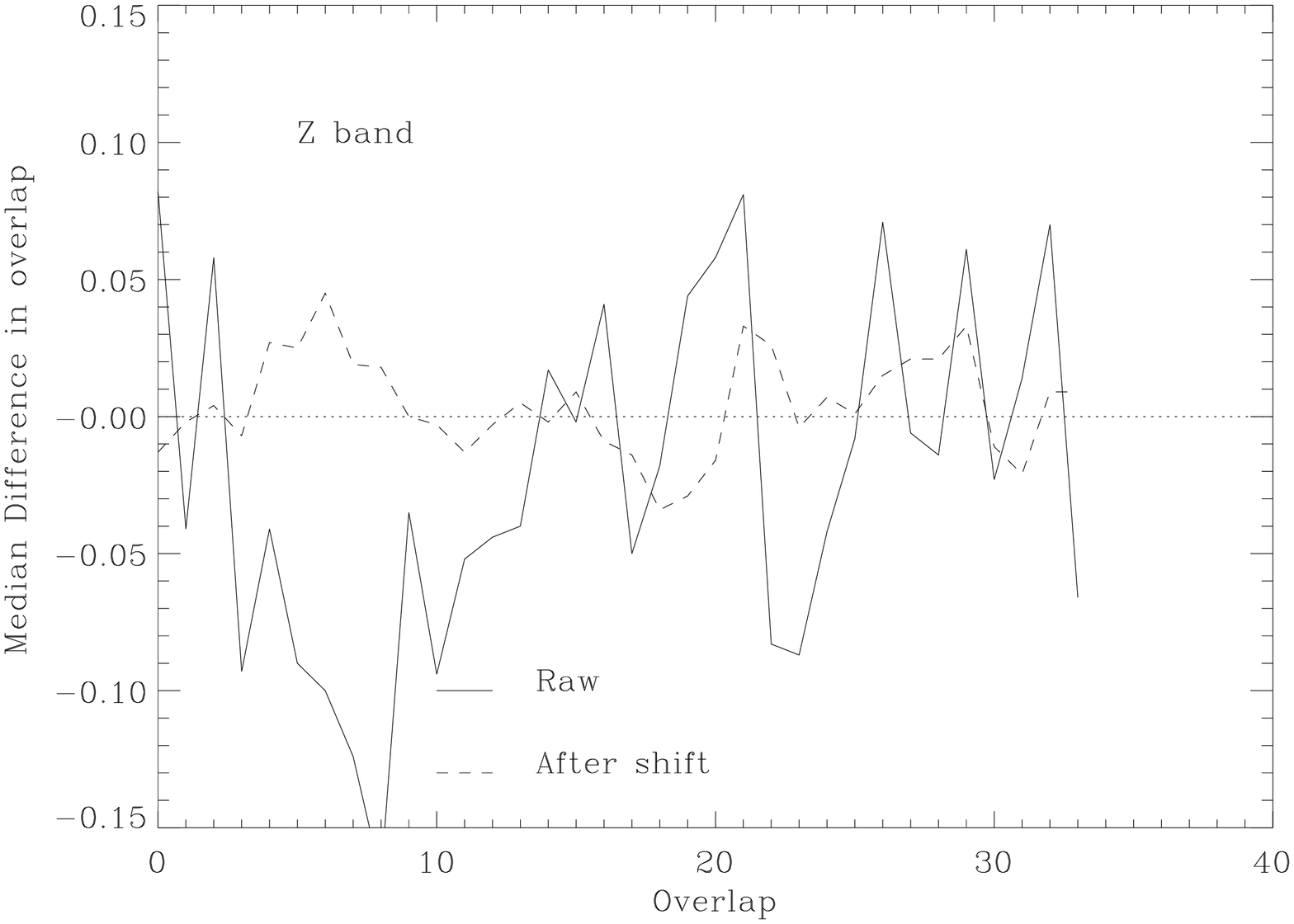}
  \caption[]{The resulting median difference in $I$-band (top panel) and $z$-band
  (bottom panel) between the 34 overlapping regions of IC\,4665. The full line indicates the
    difference in photometry before the internal calibration (``raw''), the dotted line
    after applying a photometric shift. This shift has been derived from the
    overlapping regions, putting the photometry on one internal system.}
  \label{shiftres}
\end{figure}

Some residual zeropoint offsets between the CCDs of the CFH12K mosaic were found to
be present after cross correlating stars located in the overlapping regions (see
Fig.\,\ref{spat}).  In Fig.\,\ref{fshift} we show an example of the difference in
instrumental magnitudes for the same stars observed in the overlapping region
between CCD\,06 of field\,I and CCD\,11 of field\,J (30s exposure).  The upper
part of the figure indicates the difference in I and $z$ magnitude (first and second
panel) and the considerable difference in the resulting $I$-($I-z$) CMD (third
panel). For every overlap the photometric shift in I and $z$-band was determined 
on a field-by-field and a CCD-by-CCD basis. In addition, an instrumental
magnitude difference between the various exposures times was also found
to occur. In this case we cross-correlated the stars taken at different
exposure times to obtain the photometric shifts between 2s, 30s and 300s.  In
this way an internal calibration of the full dataset was achieved. 
The lower part of Fig.\,\ref{fshift} exemplifies the improved photometric
 correspondence between CCD\,06 of field\,I and CCD\,11
of field\,J after the internal calibration. Fig.\,\ref{shiftres} shows the median difference in
$I$-band and $z$-band for all the 33 overlapping regions that were used in the internal
photometric calibration. It shows the decreased differences between the
photometry of the various overlaps.

\subsection{Infrared photometry}
\label{irphot}
The next step in the derivation of IC\,4665 properties after the optical
selection of the new candidate members using the calibrated $I$,$z$ colour-magnitude
diagram (see Sect.\,\ref{newm}) is the step of weeding out interloper
contaminants to the selected member dataset. 
$K^{\prime}$-band imaging of 101 optically selected IC4665 candidate members was obtained
with the 1k$\times$1k CFHT IR camera (Starr et
al. 2000\nocite{2000SPIE.4008..999S}) on July 10-12, 2003, and on May 30-June 3,
2004. The targets stars were chosen such as to probe the radial contamination
factor, i.e. the stars observed with CFHT-IR were selected from either field\,A,
E or L. All the observed fields with CFHT-IR are indicated in
Fig.\,\ref{spat}. The exposure time varied from 4 to 14 minutes for candidates
with I magnitudes ranging from 17 to 22. Each object was dithered on 5 to 7
positions on the detector, so that the median of the images would provide an
estimate of the sky background during the exposure. Individual images were dome
flat fielded, sky subtracted, then registered and added to yield the final
$K^{\prime}$-band image. $K$-band photometric standards from Hunt et
al. (1998)\nocite{1998AJ....115.2594H} were observed every couple of hours each
night. Aperture photometry was performed on candidates and photometric
standards. For a few close visual binaries
a large aperture was first used to estimate the system's total flux, and PSF
photometry was performed to derive the flux ratio of the system components. The
derived $K$-band photometric zero point is 22.86 $\pm$ 0.03 mag for both runs,
assuming an average extinction coefficient of 0.07 mag/airmass.

For the brighter optically selected candidate members, infrared data were
obtained from cross-correlating all selected candidate members with the
Two-Micron All-Sky Survey (2MASS) catalogues. The 2MASS catalogues are complete
down to $K_{s} \sim 14.3^{m}$ with an uncertainty of $0.05^{\rm m}$. The
completeness limit corresponds to a mass of $\rm M\sim0.20\,M_{\odot}$ at the
estimated age and distance of IC\,4665.

\section{IC\,4665 cluster member selection}
\label{newm}
\subsection{Binary sequence}
\begin{figure}
  \includegraphics[height=8cm,width=8cm]{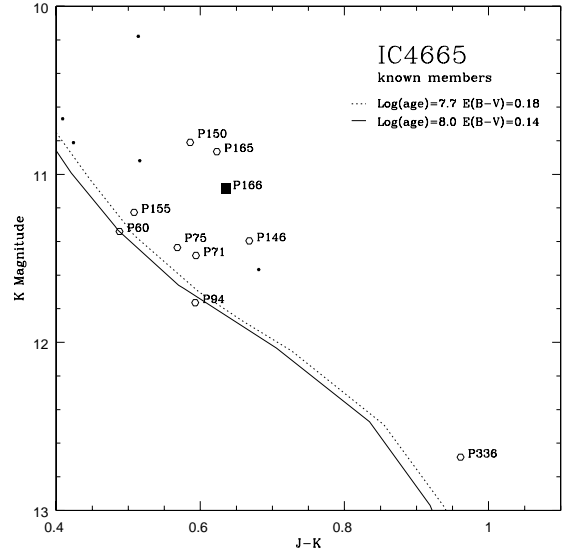}
  \caption[]{The NIR CMD for IC\,4665 members with confident $I$,$z$ photometric
    measurements (large symbols), based on 2MASS photometry.
    Small symbols are member stars without accurate $I,z$ magnitudes. Star
    P\,166 is an SB2. The figure shows that 5 out of 10 objects are candidate
    binary stars.} 
  \label{binar}
\end{figure}

The photometric selection of new members relies on the accurate positioning of
the isochrone  corresponding to the age and distance of
IC\,4665 in the optical $I$,$z$ CMD. Throughout the paper we use evolutionary tracks 
calculated by Baraffe et al. (1998)\nocite{1998A&A...337..403B} and Chabrier et
al. (2000)\nocite{2000ApJ...542..464C}  (the NEXTGEN and DUSTY models)
unless otherwise stated. In this subsection we use a total of 10 known members
(see Fig.\,\ref{prevmem}) that have reliable I and z photometry in order to
provide the actual positioning of the isochrone for an efficient member
selection; the remaining IC\,4665 members are saturated in the I and z images. A close look at the 10 
faintest IC\,4665 members in a NIR CMD (Fig.\,\ref{binar})
using the Two Micron All-Sky Survey data indicate that their distribution is split
in two groups. One half of the ten member stars is thus suspected to be binary
objects. The reality of the binary sequence is supported by the star P\,166
(filled square), a double-line spectroscopic binary (Prosser \& Giampapa
1994). Surprisingly, a binary sequence is not obvious from the optical CMD presented
in Fig.\,\ref{prevmem}, where we have indicated the names of 10 members in
question. The cluster isochrone for the selection of new members which is
presented in the next subsection is thus positioned in the $I$,$z$ CMD on the
presumably five non-binary members. 

\begin{figure*}
  \center
  \includegraphics[height=12cm,width=12cm]{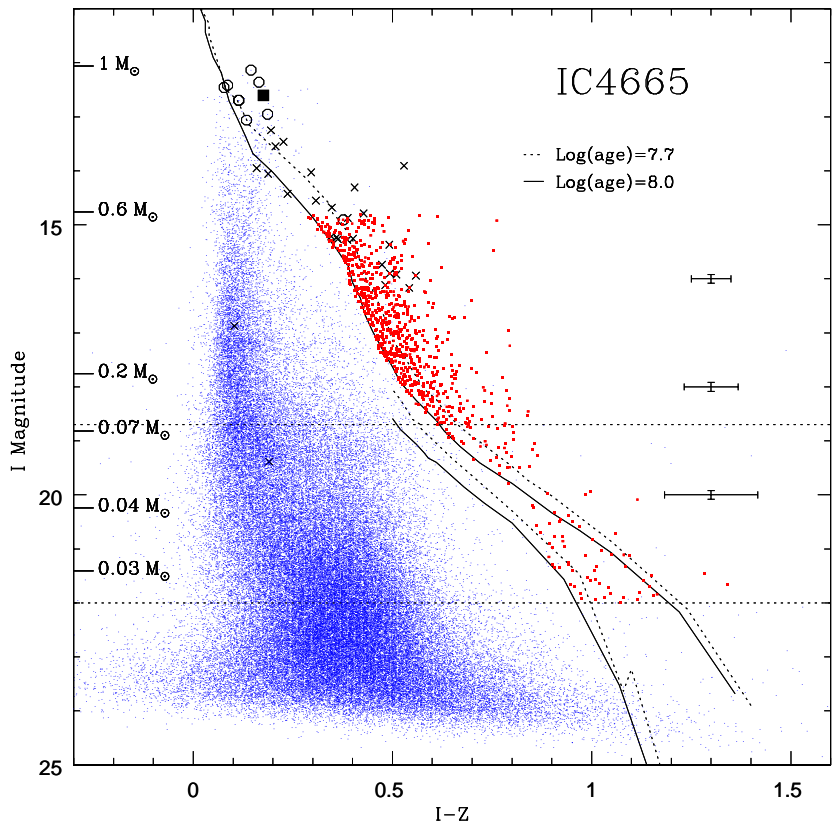}
  \caption[]{Colour magnitude diagram of CFHT $I$,$z$ observations for
  IC\,4665. Some 94
    Brown Dwarf candidate members and 691 low mass stellar candidate members
    have been photometrically selected using the theoretical isochrone
    corresponding to 100\,Myr, down to a magnitude of $\rm I=22^{m}$. Member
    stars are given as circles or as a square (the confirmed binary
    P\,166). Probable member stars identified by P92 are given as small
    crosses. Error bars have been determined from the overlap regions, with
    $\rm Y(pixel)>300$. Note that for clarity the Galactic field is plotted
    corresponding to two CFH12K fields, whereas all new candidate members are
    represented. The mass scale corresponds to the 50\,Myr NEXTGEN model for
    $>0.2\,M_{\odot}$ and to the 50\,Myr DUSTY model for lower masses.} 
  \label{newmem}
\end{figure*}

\subsection{Newly selected candidates}
In Fig.\,\ref{newmem} we present the isochronal selection of new candidate
members of IC\,4665. For reference, brown dwarfs have an expected I magnitude of
18.8 or dimmer at the estimated age and distance of the cluster. Candidate
members are selected when their $I$,$z$ photometry locates them to the right of the
100\,Myr isochrone (full line) down to a magnitude of $ I=22^{m}$, if their
PSF profile was found to be stellar. This age is a conservative estimate and in
accordance with the isochronal age of the more massive members as presented in
Fig.\,\ref{prevmem}. The upper magnitude limit ensures the completeness of
selected members down to an equivalent mass of $\rm \sim30\,M_{Jup}$. The
selection strategy therefore ensures the inclusion of all members, given the
uncertainties. The ten member stars described in the previous paragraph are
depicted by the large symbols at the top of the figure. The 100\,Myr isochrone
is positioned to go through the bluest confirmed members of the single star
sequence. The isochrone actually envelopes also quite well the probable members
which are indicated by the small crosses. The presence of a binary sequence is
also confirmed by the CFH12K $I$,$z$ photometry.

Fig.\,\ref{newmem} in fact shows the 100\,Myr isochrones (full lines) of
the NEXTGEN models (Baraffe et al. 1998) and DUSTY models
(Chabrier et al. 2000). For reference the 50\,Myr (dotted lines) isochrones are
drawn as well. DUSTY model calculations include the settling of dust in the
upper atmospheres of cool objects, which is not included in the NEXTGEN models.
The transition between the dusty atmosphere and gaseous atmosphere is thought to
occur at $ (I-z)\simeq0.85$ corresponding to a NEXTGEN $T_{\rm eff}$ of
3000\,K. This is the reason for the artificial ``hook''  in the selected objects
at this colour.  We especially note that had we chosen {\it only} NEXTGEN
models for selecting candidate members of the lowest mass and not the DUSTY
models, then only a very small amount of BDs candidates would have been
found. 

At the distance of IC\,4665, the 100\,Myr isochrone intersects the foreground
and background main sequence and giant branch stars in the CMD at about $\rm
0.8\,M_{\odot}$, rendering the selection procedure at these masses highly
inefficient. For clarity, Fig.\,\ref{newmem} does not show all the stars
extracted from the three exposures, but for two fields only (A and E). This
translates to about one sixth of the total number of extracted stars in the
surveyed area. On the contrary, all the selected candidate members are plotted.
The selected objects were further checked by eye to eliminate any spurious
detection due to hot pixels and/or bad CCD columns. The selection procedure
delivered a total of 94 brown dwarf candidate members ($ 18.8^{m}<I<22^{m}$)
from 300s measurements, 529 low mass stars ($15.8^{m}<I<18.8^{m}$) measured in
the 30s exposure and 163 low mass stars ($14.8^{m}<I<15.8^{m}$) measured in the
2s exposures. A list of these objects is found in Table\,3 (electronic version
only).

\section{Contamination by foreground and background stars}
\label{cont}
The contamination in the foreground by faint M-dwarfs and by highly reddened
bright objects behind the cluster in the direction of IC\,4665 is anticipated to
be quite substantial. The large numbers of newly selected candidate members
makes this obvious. In this section we try to quantify the statistical
contamination fraction and constrain memberships using $K$-band photometry and
public proper motion data.

\subsection{Statistical contamination}
\label{bgc}
An estimate for the statistical contamination is derived from the two control
fields. These fields are located well beyond the cluster boundary of IC\,4665.
The same selection procedure to identify candidate members as explained in the
previous section was applied to the two control fields. In the low mass stellar
range down to the hydrogen burning limit ($14.8 < I <18.8$), we select 108
(48+60) objects in the two control fields. This translates to an absolute
contamination of $164\pm13$ stars per square degree. The surveyed area
centred on IC\,4665 equals effectively $3.82$ square degrees, and we thus
expect $626$ contaminating stars. We selected $691$ IC\,4665 candidate members and
we thus find a statistical contamination fraction of $\sim 85\%$. That is only 1
in 6 photometrically selected candidate member is expected to be an actual
cluster member. We note that contamination is expected to increase for
the brighter stars, as IC\,4665 sequence intersects the Galactic field giant
branch at $I<15^{m}$. For example in the range ($14.8 < I < 15.8$), i.e. the
candidate members selected from the short exposure, we calculate 90\%
contamination. In the substellar domain, corresponding here to $18.8 < I <
22$, we estimated the background contamination from the 300s exposures. In the
control fields (9+2) 11 objects were selected from the 300s exposure, i.e. $\rm 17\pm5$ per square
degree. Therefore we expect to find $\sim 64$ contaminants. In the surveyed area
a total of 94 substellar candidates are found in the 300s exposures. We conclude that the statistical
contamination of our photometric selection of BD candidate members is on the
order of 70\%. A more sophisticated estimate of the fractional contamination as
function of $I$-band magnitude bins is not reliable. The cluster signal does not
reach a high enough S/N per bin. 

The relatively low efficiency in selecting new member object compared to the
efficiency reported by Moraux et al. (2003) for a similar procedure of
photometric selection for new \object{Pleiades} members can be attributed to a
smaller (absolute) Galactic latitude ($b=-23$ for Pleiades, $b=+17$ for
IC\,4665). A second factor yielding a much higher fractional contamination is the
larger distance to IC4665 (350\,pc against
125\,pc) resulting in $2.2^m$ difference in apparent magnitude. The IC4665
stellar and substellar sequences are therefore much closer in magnitude to the
bulk of the foreground and background stars in a CMD. This difference is only partially
alleviated by the younger age of IC\,4665, as can be judged from the location of
the isochrones corresponding to 50\,Myr and 100\,Myr in Fig.\,\ref{newmem}.

\begin{figure}
  \includegraphics[height=8cm,width=8cm]{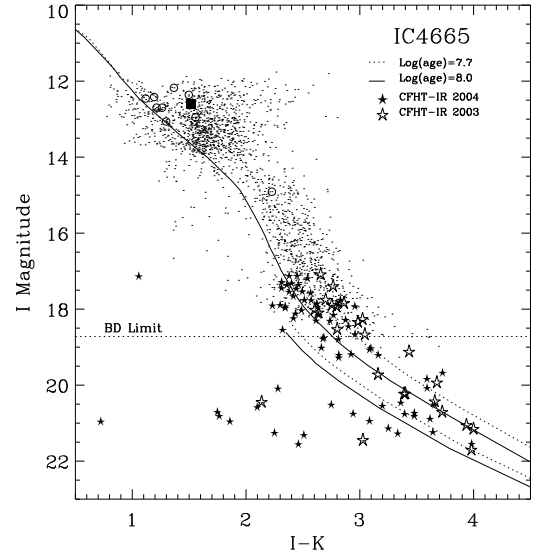}
  \caption[]{Separating contaminant stars from optically selected objects using $K$-band
    photometry, either from the 2MASS survey (small dots) or observations with CFHT-IR instrument.}
  \label{nircmd}
\end{figure}

\subsection{Contamination estimate from $K$-band photometry}
\label{Kphot}
A first step in actually removing contaminant objects is by using $K$-band
photometry. Although from first principles cluster BDs and stars are expected to
be only intrinsically more luminous than field stars of the same effective temperature
(Zapatero Osorio et al. 1997\nocite{1997A&A...323..105Z}), IR photometry does effectively separate
out a fraction of non-member objects as shown in the case of $\alpha$\,Per
(Barrado y Navasqu\'{e}s et al. 2001\nocite{2001ApJ...546.1006B}). In part the explanation may be
sought in photospheric differences, not large enough to produce a significantly
different ($I-z$) colour, but that do become significant in the ($I-K$) colour. This
is exemplified in the optical-NIR CMD of Fig.\,\ref{nircmd}.  The small dots and
open/filled asterisks represent all the selected stars based on the ($I-z$) CMD,
for which $K$-band photometry is available, observed by either 2MASS or with
CFHT-IR as described in Sect.\,\ref{irphot}.  A fraction of the ($I-z$) colour
selected objects are separated out in the diagram, not being located near the
appropriate isochrone. The $K$-band contamination as function of $I$-band magnitude
displays two peaks of $\sim35\%$ contamination at $I$-band magnitude of $14^{m}$ and
$18^{m}$, i.e. the loci where the isochrones intersect the bulk of the field
population. 

The correction step using $K$-band on the optically selection procedure has a
relatively low efficiency for stars brighter than $I\sim18^{m}$. This can be
judged from the 2MASS data in the NIR CMD of Fig.\,\ref{nircmd}. At these
magnitudes the $I-K$ colour allows at maximum the elimination of 35\% per
$I$-magnitude bin. A better estimate of the efficiency is to perform the same
analysis using the control fields. Cross-correlating the optically selected
objects in the two control fields (surely non-members) with the 2MASS dataset
then we find that 80\% of these objects are to the right of the cluster
isochrone down to the $I\sim18^{m}$ in a $I$,($I-K$) CMD, and thus consistent with
membership. For magnitudes fainter than $I\sim18^{m}$ using the CFHT-IR data,
Fig.\,\ref{nircmd} shows that $K$-band photometry has an efficiency of $\sim35\%$. 
We see thus that $K$-band photometry does a reasonable job in separating out
probable non-member stars from members.
However the lower fractional contamination as determined from the
$K$-band compared with the statistical estimate of the previous subsection
directly implies that $K$-band is not the definitive answer to construct a clean
sample of cluster members, at least not for the age, latitude and distance of
IC\,4665. Additional NIR photometry at other wavelength bands does certainly
allow to constrain membership of potential candidate members that have initially
been selected by making use of isochronal colours and magnitudes, as we'll see in
the following section.

\begin{figure}
  \includegraphics[height=7cm,width=8cm]{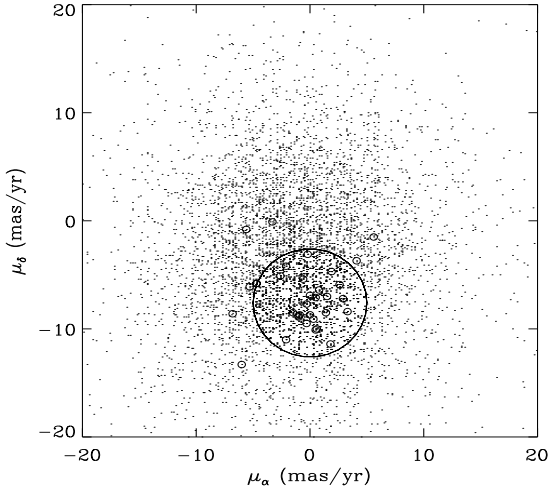}
  \caption[]{A vector point diagram for stars within 1 degree from IC\,4665,
    based on UCAC2 (Zacherias et al. 2004). The estimated bulk proper motion (circle) for
    IC\,4665 based on the brightest stars using Tycho-2 data.}
  \label{pm1}  
\end{figure}

\begin{figure}
  \includegraphics[height=7cm,width=8cm]{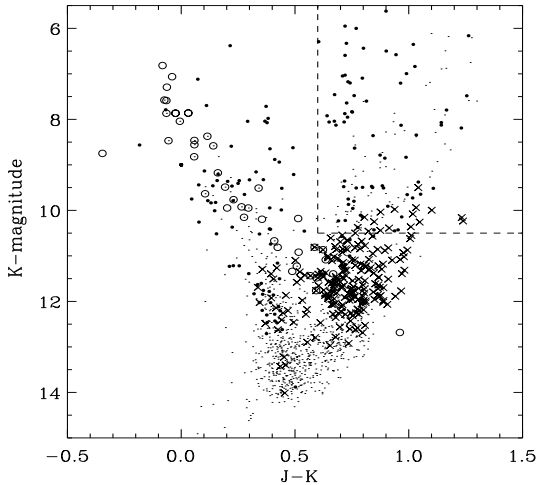}
  \caption[]{The colours and magnitude for stars with proper motion consistent
    with IC\,4665 (small dots), i.e. that are inside the circle of
    Fig.\,\ref{pm1}). Confirmed member stars (small circles) are added for
    reference. Small filled symbols are P93 non-members. Crosses are objects with UCAC2 proper motion that have
    been selected as candidate members in the optical survey.}
  \label{pm2}
\end{figure}

\subsection{Membership constraints from proper motion surveys}
\label{pm}
The proper motion of IC\,4665 is small and does not stand out significantly from
the bulk field motion to be a strong constraint on membership. Hipparcos
measured the astrometry of 13 IC\,4665 members and obtained the following proper
motion in mas\,yr$^{-1}$: $\mu_{l*}=-7.2\pm0.3$ $\mu_{b}=-3.0\pm0.3$ or
converted to equatorial coordinates
$(\mu_{\alpha},\mu_{\delta})=(-0.4,-7.5)$\,mas\,yr$^{-1}$ (Hoogerwerf et
al. 2001\nocite{2001A&A...365...49H}). Nevertheless, here we explore two public
proper motion catalogues to evaluate membership for the optically selected
candidate members, {\it viz.} the Tycho-2 catalogue (H\o g et
al. 2000\nocite{2000A&A...355L..27H}), and the USNO UCAC2 (Zacharias et
al. 2004\nocite{2004AJ....127.3043Z}). The Tycho-2 catalogue is complete down to
$K_{s}=8^{m}$, and provides proper motion with an accuracy of
2.5\,mas\,yr$^{-1}$. UCAC2 on the other hand is not complete for stars
brighter than $R\simeq8^{m}$ due to saturation problems. For the stars in the
R-band magnitude interval $13-16$, UCAC2 provides proper motion with an error of
6\,mas\,yr$^{-1}$. Cross-correlation with 2MASS All Sky Catalogue learns that
97\% of the 2MASS entries are in the UCAC2 catalogues for $K$-magnitude between $8-11$.

We find 35 IC\,4665 members in the UCAC2 catalogue for a mean proper motion of
$(\mu_{\alpha},\mu_{\delta})=(0.0,-7.6)$\,mas\,yr$^{-1}$ with a 0.8 error in the
mean value and a $\sim 5$\,mas\,yr$^{-1}$ standard deviation in both the right
ascension and declination proper motion distributions. The UCAC2 proper motion
for IC\,4665 is in good correspondence with the Hipparcos result of Hoogerwerf
et al. The spread in proper motion space, i.e. the internal motion of IC\,4665
can be judged from Fig.\,\ref{pm1}, in which the 35 member stars are indicated
by the small open circles. Some 14 of these members are also found in
Tycho-2. Their Tycho proper motions deviate from the UCAC2 results with
$(\delta\mu_{\alpha},\delta\mu_{\delta})=(3,3)$\,mas\,yr$^{-1}$, consistent
within the uncertainties. The only discrepant result between the two catalogues
we find for the visual binary K\,67, whose UCAC2 proper motion is not consistent
with membership, but its Tycho-2 is.

Using UCAC2 to identify proper motion members to IC\,4665, we require for
selection that a star has a proper motion that lies within 5\,mas\,yr$^{-1}$ of
the cluster proper motion. This threshold is indicated by the large circle in
Fig.\,\ref{pm1}. It equals the standard deviation in the UCAC2 proper motion of
the member stars and this criteria will therefore exclude 32\% of genuine
cluster members. It is a compromise between optimising member selection and
excluding background objects. We stress that proper motions deviant from the
internal motion of the cluster as defined above is not a strong enough
constraint to exclude membership.

In Fig.\,\ref{pm2} we present the result of the proper motion selection in a
CMD. The tiny dots are the proper motion selected candidate stars that are
within the encircled area of the vector point diagram in Fig.\,\ref{pm1}.  The
cluster sequence is clearly recovered, and {\it all} member stars (open circles)
have been plotted to indicate this. The small filled circles indicate all the
rejected objects from the P93 catalogues. Note that on and near the cluster
sequence for stars with a $K$-magnitude brighter than 10.5 (corresponding to
about $1.2\,M_{\odot}$) objects are located that do not have a counterpart in
the P93 catalogues within a $2\arcsec$ cross-correlation radius, neither as a
rejected nor as a confirmed member. Some among them are located within
$1^{\circ}$ of the cluster centre, i.e. within the tidal radius of the
cluster. These objects are clearly candidate IC\,4665 members. The crosses in
Fig.\,\ref{pm2} identify the optically selected candidate members for which a
UCAC2 proper motion exists. Fig.\,\ref{pm2} shows that currently public proper
motion information for IC\,4665 is limited to members brighter than
$K\simeq12^{m}$ or equivalently to about $\rm 0.6\,M_{\odot}$.  The $(J-K)$
colour provides a strong constraint for cluster membership in this case. Deep
NIR observations in addition to the $K$-band are therefore strongly
warranted. The new generation of NIR wide field cameras (WFCAM, WIRCAM) are
especially designed to perform large surveys of young open cluster, e.g. like
the Galactic Cluster Survey as part of the UKIDSS survey. For now we consider
the stars that are too red in the upper of the diagram at color $J-K>0.6$ and
$K<10.5$ as likely non-members, and that the highest mass cluster stars are
still on the main sequence. All eleven red giants except one that are brighter
than $K<5.5$ and lie outside the plotted range in Fig.\,\ref{pm2} are
confirmed non-members.

\section{Discussion}
\label{disc}
The radial and the mass distribution for members and candidate members are
derived and discussed.

 \begin{figure}
   \center
   \includegraphics[height=7.5cm,width=8cm]{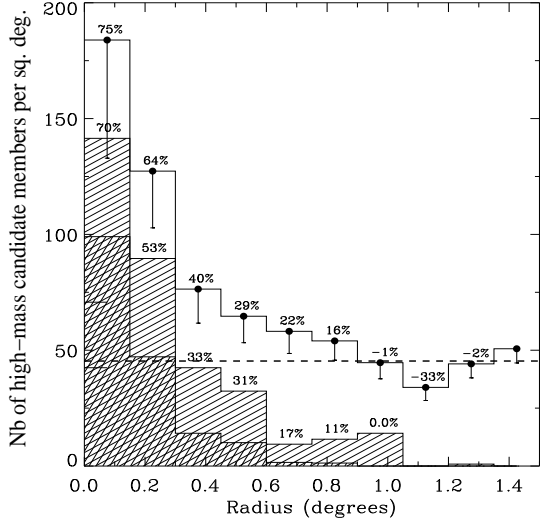}
   \caption[]{Histogram for stars with $K<10.5$ and $J-K<0.6$. Plain
   histogram 2MASS All Sky Catalogue selected stars; single-hatched histogram all
   selected stars from Prosser catalogues; double-hatched histogram confirmed
   members. Dashed line is the average surface density of control field C1 and C2. Upper ratios:
   2MASS to background. Lower ratios: selected stars to confirmed members.}
   \label{radmem}
 \end{figure}

 \begin{figure*}
   \center
   \includegraphics[height=6.5cm,width=6cm]{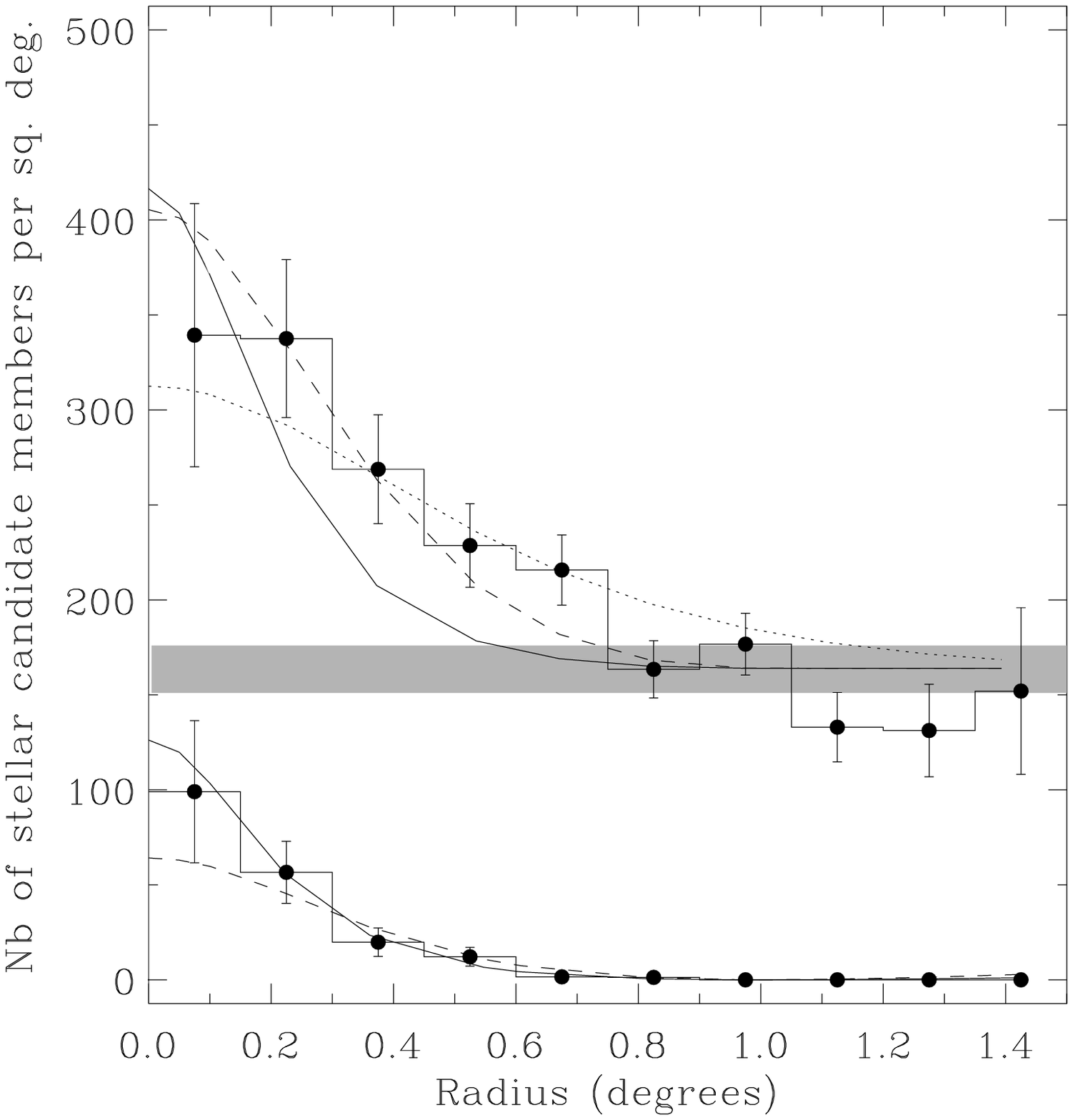}
   \includegraphics[height=6.5cm,width=6cm]{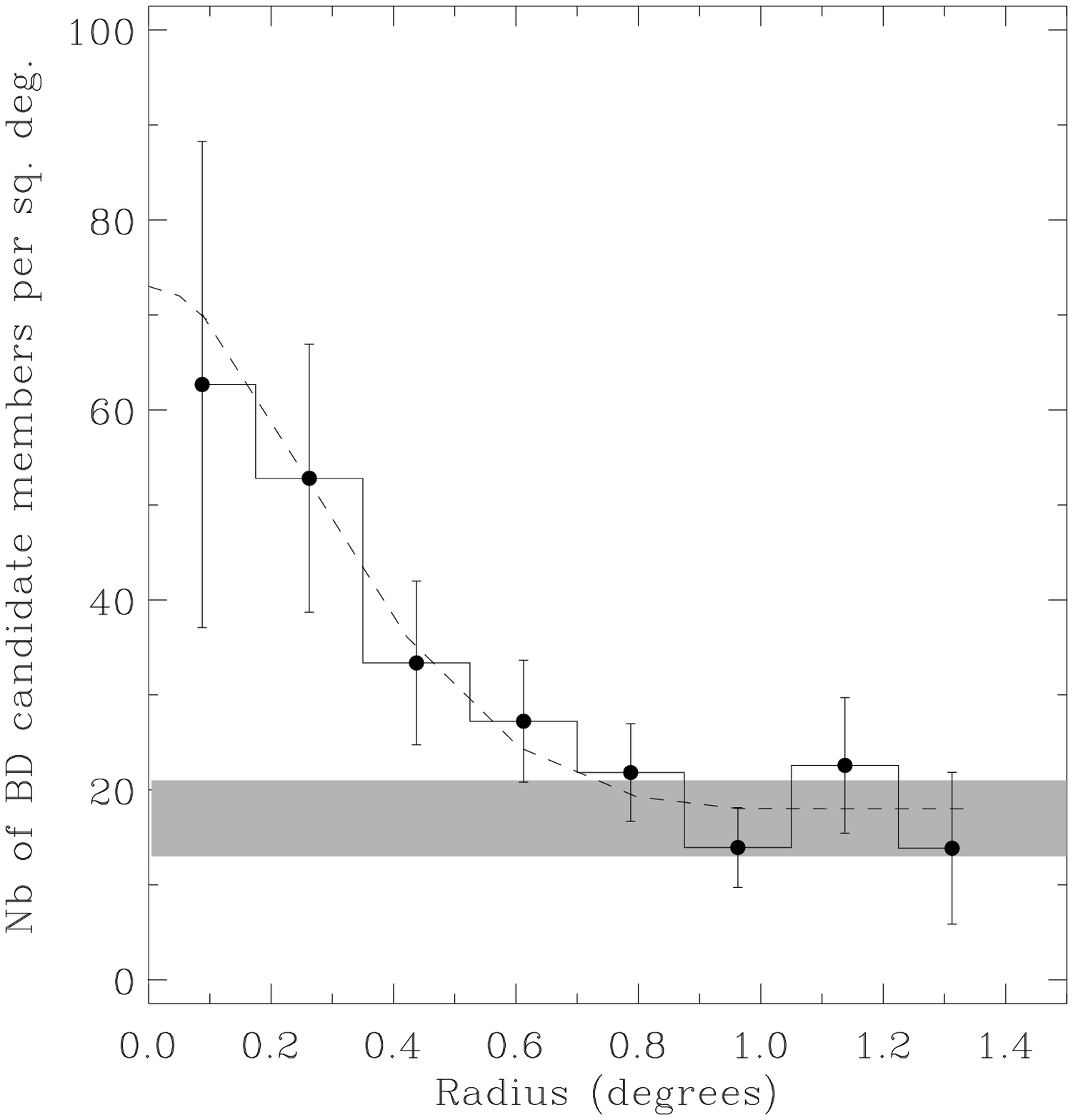}
   \caption[]{{\it Left:} Upper histogram presents the radial distribution for
   candidate members with  
   $14.8<I<18.8$. Lower histogram equals the double-hatched histogram of Fig.\,\ref{radmem}.
   The shaded areas are the stellar background estimates
   derived from the control fields. Full lines are King profiles with $\rm
   r_{c}=0.25^{\circ}$ and $\rm r_{t}=1.0^{\circ}$; dashed lines King profiles with
   $\rm r_{c}=0.5^{\circ}$ and $\rm r_{t}=1.0^{\circ}$; dotted line is the scaled
   Pleiades low-mass profile with $\rm r_{c}=0.71^{\circ}$ and $\rm
   r_{t}=1.97^{\circ}$.
   {\it Right:} The radial distribution for the BD candidate members.}
   \label{radstar}
 \end{figure*}

\subsection{The radial distribution for members and candidate members}
\label{raddis}
The radial dependence of the surface density in square degrees is derived for
the high-mass member stars and the optically selected low-mass candidate members. For the
latter we distinguish between stellar and BD candidates.  To obtain
these distributions, we first determine the mass-weighted central coordinates of IC\,4665 using
the high-mass members only. Adopting weights equaling the third power of the flux in the $V$-band
magnitude, the cluster is found centred on $\rm RA(2000)= 17^{h}46^{m}04^{s}$, $\rm Dec(2000)=
+05^{\circ}38\arcmin53\arcsec$ .

Three radial distributions for high-mass members are presented in 
Fig.\,\ref{radmem}. High-mass stars in this case are stars with a K-band
luminosity smaller than $K=10.5$ or equivalently a mass larger than
$\sim1.2\,M_{\odot}$. This mass threshold corresponds to the location in a CMD
where the cluster sequence intersects the background population
(e.g. Fig.\,\ref{newmem}). We estimate the completeness of the cluster census as
function of radius and the total radial extent of IC\,4665 using both the P93
members and the unbiased 2MASS catalogues. From 2MASS we select high-mass stars
located within $1.5^{\circ}$ of the centre that have a colour bluer than
$J-K=0.6$.  The colour requirement excludes the foreground red GB and AGB stars,
a population that makes up a large contaminant fraction at $K$-band luminosities
$K<10.5$ (see Fig.\,\ref{pm2}). The plain histogram with error bars in Fig.\,\ref{radmem} depicts the 2MASS selected high-mass stars. The
radial extent of IC\,4665 is easily read off as the distribution is seen to
attain background levels at radii larger than $\sim1^{\circ}$. The dashed line
denotes in fact the average surface density determined for a similar sized area
centred on the C1 and C2 control fields. The control field surface density
clearly marks the level of the stellar background level at the coordinates of
IC\,4665.

For comparison we added to Fig.\,\ref{radmem} two radial
distributions of P93 member stars: a single-hatched and a double-hatched
histogram.  They represent the surface density profile for all candidate members
with $K<10.5$ and $J-K<0.6$ and the census of confirmed high-mass members with
the same colour and $K$-band magnitude constraints (29 objects),
respectively. The candidate high-mass members have been investigated for cluster
membership using various criteria in various studies. Among them there exist
confirmed members, confirmed non-members and stars without definite
classification (see Sect.\,1). The bin at $1^{\circ}$ of the single hatched
histogram contains zero confirmed candidates. This confirms the maximum extent
of the cluster derived from 2MASS selected stars.

We note the ratios written above the bins of the plain and single-hatched
histograms. The ratios of the plain histogram indicate the statistical fraction
of cluster stars as function of radius, given the background level (dashed
horizontal line). The ratios on the single-hatched histogram indicate the number
of confirmed high-mass members to high-mass candidates. The two ratios at each
bin are in good agreement, indicating that the current knowledge of the {\it
relative} number of high-mass member stars of each bin is a good representation
of the actual relative number. If the radial distribution of cluster members is
best-described by a tidally truncated King profile (King 1962\nocite{1962AJ.....67..471K}), we can
thus fit and compare the profile of high-mass members with that of the low-mass
constituency adopting a tidal radius of $1^{\circ}$. The bottom histogram of the
left panel in Fig.\,\ref{radstar} shows again the distribution for the
high-mass members, i.e. the double-hatched histogram of Fig.\,\ref{radmem}. A by-eye
fit to the first four radial bins points to a core radius $r_{c}$ equaling
$\sim0.25^{\circ}$ (full line). In order to illustrate the uncertainty on this
core radius, a King profile for $r_{c}=0.5^{\circ}$ is also indicated by a
dashed line. This profile still fits the observed distribution within the error
bars. We consider it as the upper limit to the high-mass star core radius for
IC\,4665.  

The upper histogram of the left panel in Fig.\,\ref{radstar} shows the radial
dependence of the optically selected candidate stellar members in the mass range
of $0.07$ and $0.6\,M_{\odot}$. Error bars are equivalent to the statistical uncertainty for each bin. 
We note first that the surface densities derived for the control fields
(shaded areas) are in good agreement with surface densities at radii larger than
$r_t=1.0^{\circ}$. This is the case for both the low-mass stars and the BD
candadite members. It shows that IC\,4665 cannot be much more extended on the
sky, which gives confidence that the area covered with the CFH12K camera has
been sufficient (see Fig.\,\ref{spat}). The King profiles derived for the
high-mass stars are scaled and plotted using the same line styles. A comparison
between the best-fit high-mass King profile (full line) with the radial
distribution of low-mass stars reveals that the two are not consistent. On the
other hand a King profile with $r_{c}=0.5^{\circ}$ (dashed line) is in reasonable
correspondence. We thus see that the low-mass stellar population of IC\,4665 has
a $r_{c}\ga0.5^{\circ}$. One can thus conclude that for $r_{c} =0.5^{\circ}$ the
surface densities of high-mass and low-mass stars are consistent within the
statistical uncertainties. Nevertheless, it is probable that the low-mass stars
have a broader distribution on the sky than the high-mass stars. This is
illustrated by the dotted King profile in Fig.\,\ref{radstar}. It corresponds to
the distribution for the low-mass stars of the Pleiades. Scaled to the distance
of IC\,4665 this profile is described by $r_c=0.71^{\circ}$, and
$r_t=1.97^{\circ}$ and is not extremely far off the observed distribution of the
low-mass candidate members of IC\,4665.

The right panel of Fig.\,\ref{radstar} displays the BD candidate distribution
for which only a profile with $r_c=0.5^{\circ}$ is overlaid. This profile
corresponds best to the observed distribution, the uncertainties are however
large. Nevertheless a King profile fit with $r_c$ of 0.25 (high-mass stars) or
0.71 (low mass stars in Pleiades) are not consistent with the observed
distribution (not plotted). We conclude that the radial distribution of BD
candidates is less certain than the one of the low-mass stars, however it seems
more reminiscent to the low-mass star than the high-mass star distribution. It
would indicate a mixed BD and low-mass stellar population in IC\,4665. On the
contrary the high-mass stars may occupy a smaller area on the sky, which may
constitute the first indication of present-day mass segregation in
IC\,4665. It is worth mentioning that mass segregation between stars of
masses $\rm > 3\,M_{\odot}$ and lower mass stars has been reported for the open
cluster \object{NGC\,2547} (Littlefair et al. 2003\nocite{2003MNRAS.345.1205L};
Jeffries et al. 2004). This cluster has an age that is probable only slightly
less than the one in IC\,4665, and the clusters are of comparable
mass. Littlefair et al. and Jeffries et al. provide evidence in favour of a
primordial origin for the mass segregation of NGC\,2547, adding to other
instances of inferred primordial mass segregations in case of more massive
clusters (see Bonnel \& Davies 1998\nocite{1998MNRAS.295..691B}; de Grijs et
al. 2002\nocite{2002MNRAS.331..245D}). Therefore the reality of mass segregation in IC\,4665
between stars of $\rm \gta 1.2 M_{\odot}$ and lower mass stars is of
considerable importance, but in need of confirmation.

Finally, using the King profiles for the low-mass stars and BDs candidates we can
estimate their total number in the cluster. IC\,4665 consists of $30\pm10$ BDs
and $105\pm20$ low-mass stellar members, where the uncertainties are estimated from
trying King profiles for various normalization factors. A background
contamination within the tidal radius then proves to be $80\pm5\%$ and
$60\pm10\%$ for low-mass stars and BDs respectively; within a radius of
$0.55^{\circ}$ contamination factors of $65\pm5\%$ and $40\pm15\%$ are
expected. These estimates also take into account the statistical uncertainties
in the background level. From the radial distribution we estimate the fraction
of BDs to total members in IC\,4665. At the upper mass end there are 35 members
with masses in the range $\rm 1<M/M_{\odot}<5$ and an estimated 39 stars between
1 and $\rm 0.6\,M_{\odot}$, based on a Salpeter IMF. Taking into account the
uncertainties we find a range of 10-19\% for the fraction of IC\,4665 BDs 
(down to a mass of $\sim 0.03\,M_{\odot}$) to
total members. These factions are consistent with fractions found for other
young open clusters for which such data are available, {\it viz.}  10-15\% for
the \object{Pleiades}, \object{NGC\,2516}, and \object{Blanco\,1} (see Moraux et
al. 2005\nocite{2005MmSAI..76..265M} and references therein). In this respect IC\,4665 thus seems
to be  a ``normal'' young open cluster.

%

\subsection{The mass function}

\begin{table}[t]
  {
    \begin{center}
      \caption[]{The mass function for IC\,4665 for an age of 50Myr. First
        column gives the central $I$-band magnitude of each bin. The column indicated with
        'f' is the correction factor applied to account for the contamination by
        non-member objects.}
      \begin{tabular}{ccccrrrr}
        \hline
        \hline
        $I$ & $\rm \Delta M$       & $\rm N_{obs}$ &  f    & $\rm N_{corr}$ & dN/dM & dN/dlogM \\
            & ($\rm M_{\odot}$)&                   & \%   &              &                \\
        \hline
        13.30 & 0.92-0.77   & 111 & 15 & 16.7 & 109.6 & 213.0\\
        14.30 & 0.77-0.66   &  57 & 15 &  8.6 &  75.0 & 122.9\\
        15.30 & 0.66-0.47   & 115 & 15 & 17.3 &  93.4 & 120.2\\
        16.18 & 0.47-0.31   & 107 & 20 & 21.4 & 134.5 & 119.8\\
        16.93 & 0.31-0.20   & 114 & 20 & 22.8 & 203.7 & 118.5\\
        17.68 & 0.20-0.135  &  96 & 20 & 19.2 & 291.2 & 111.1\\
        18.43 & 0.135-0.093 &  62 & 20 & 12.4 & 299.2 &  77.8\\
        19.23 & 0.093-0.059 &  25 & 40 & 10.0 & 287.6 &  49.5\\
        20.08 & 0.059-0.044 &  11 & 40 &  4.4 & 293.2 &  34.3\\
        20.93 & 0.044-0.035 &  17 & 40 &  6.8 & 799.6 &  72.3\\
        \hline
      \end{tabular}
      \label{ta:imf}
    \end{center}
  }
\end{table}

\begin{figure*} 
  \center
  \includegraphics[height=5.5cm,width=6cm]{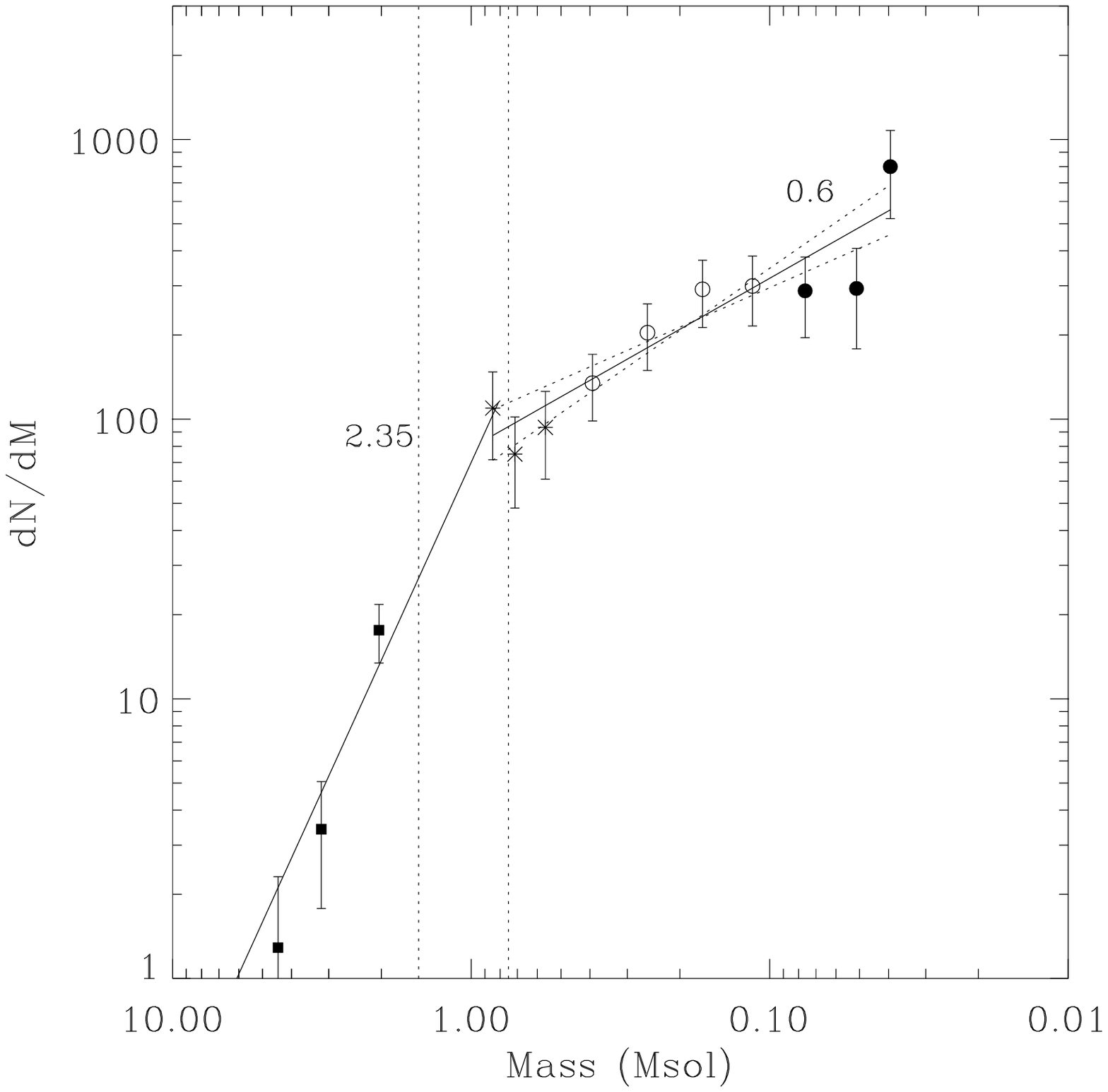}
  \includegraphics[height=5.5cm,width=6cm]{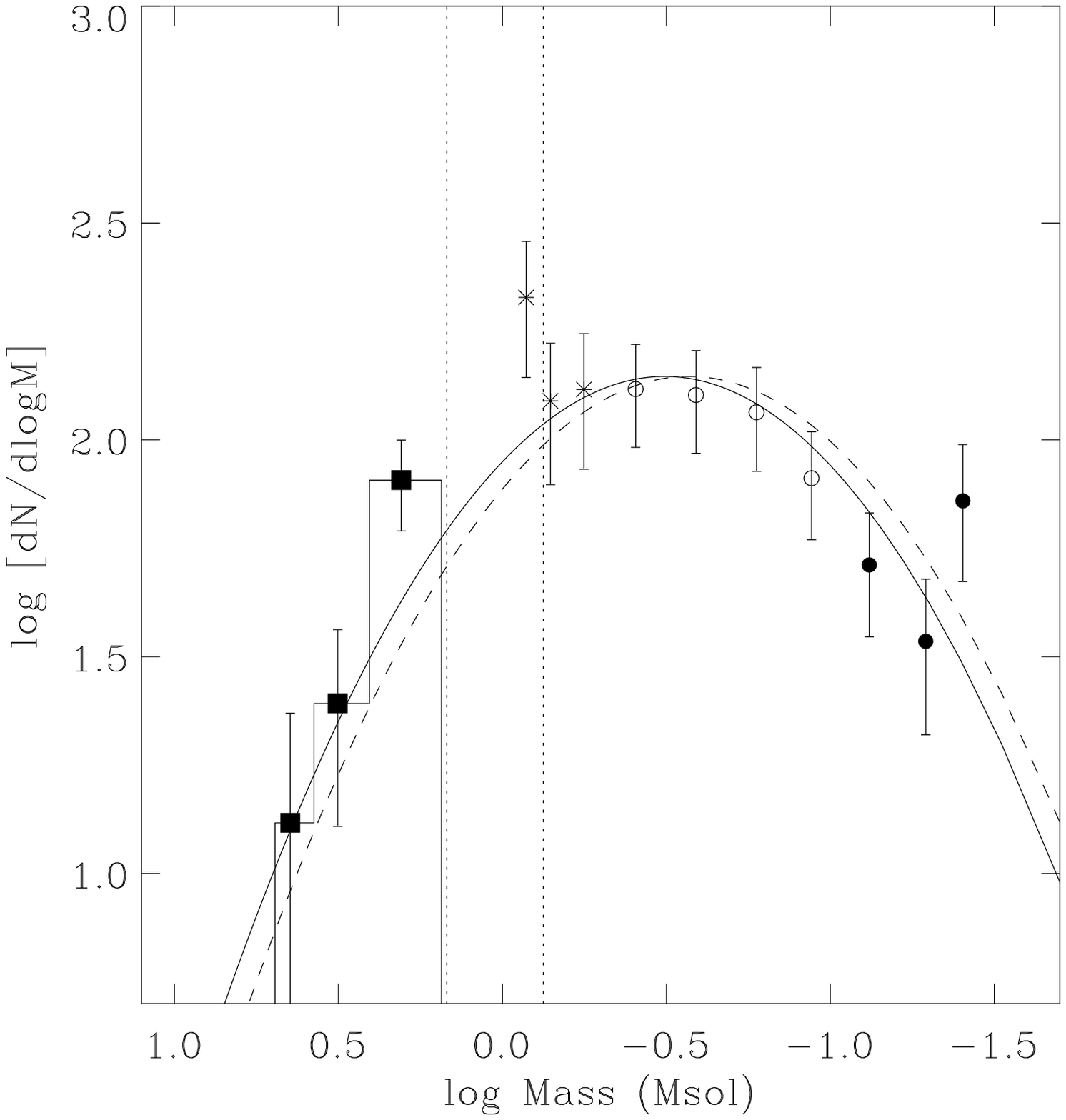}
  \includegraphics[height=5.5cm,width=6cm]{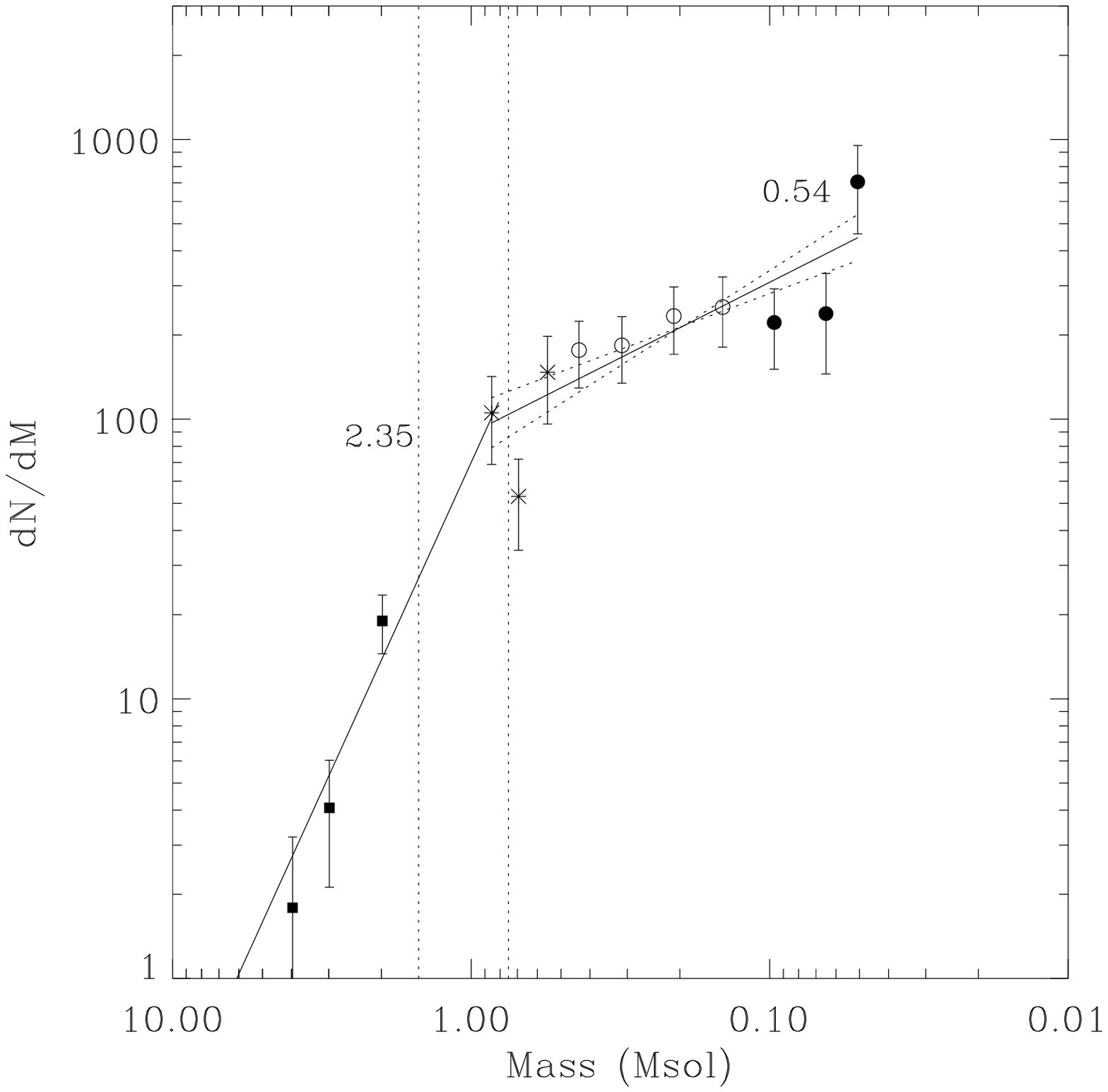}
  \includegraphics[height=5.5cm,width=6cm]{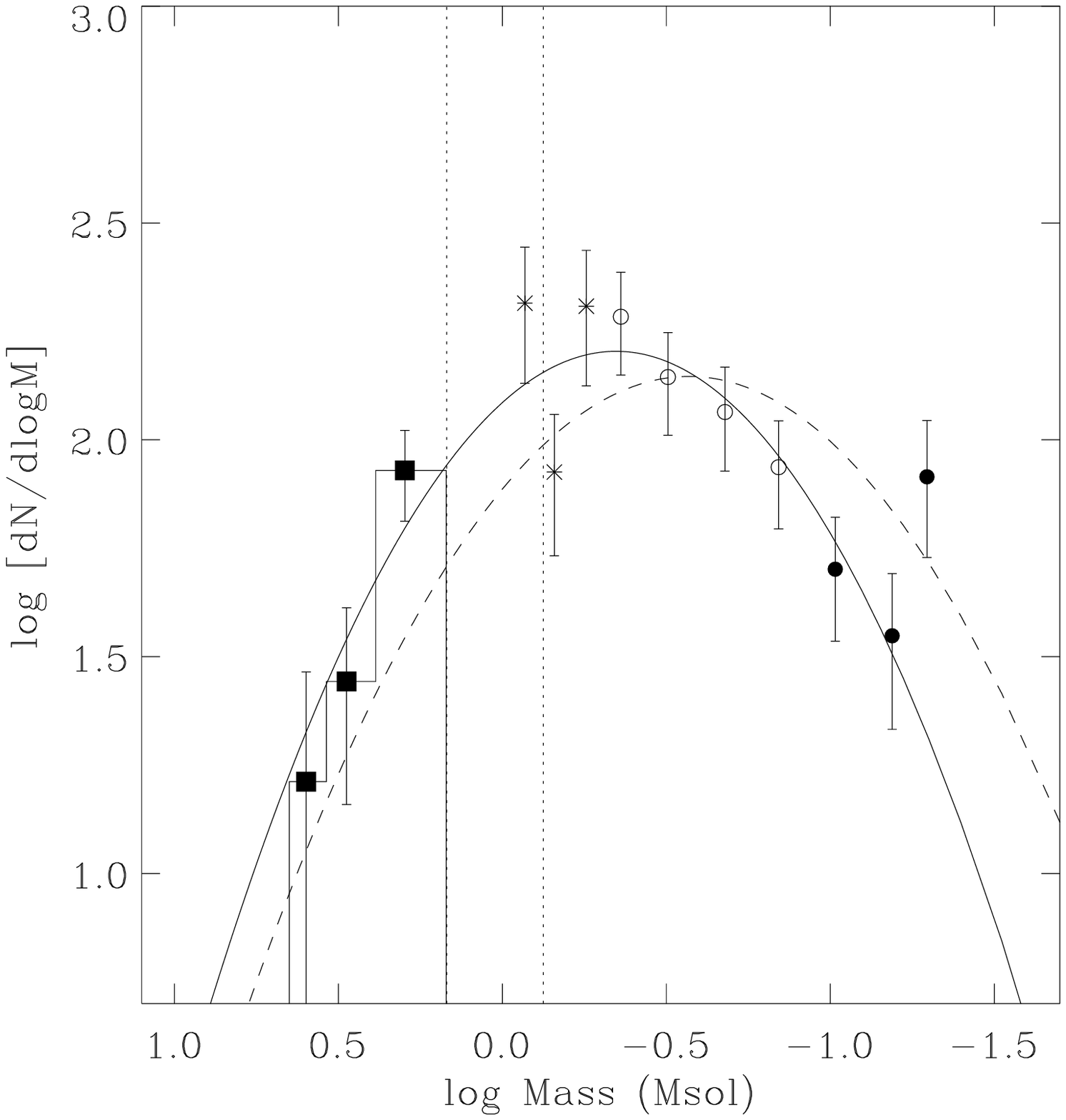}
  \caption[]{The IC\,4665 mass function in the dN/dM (left column) and
  dN/dlog(M) (right column); top row assuming an age of 50\,Myr, bottom row for
  an age of 100\,Myr. Symbols and lines in each panel have the same
  meaning. Squares denote candidate members within the tidal
  radius of $1.0^{\circ}$ selected from 2MASS, asterisks, open circles and
  filled circles from the 2s, 30s and 300s exposures. The vertical dotted lines
  indicate the masses for which the cluster sequence  intersects the background stellar
  population. Full lines in left column are Salpeter-law mass functions and power
  law functions (fits) with slopes of respectively $-0.6$ and $-0.54$, and their
  associated $1\,\sigma$ fitting uncertainties (dotted). In the right columns lognormal fits to the
  observed mass functions (full line) and the scaled Pleiades lognormal fit (dashed) 
  are indicated (see text for details). Note the differences between the
  logarithmic mass functions for the two adopted ages.}
  \label{mf}
\end{figure*}

The age of IC\,4665 is not better determined than within a margin of nearly
40\,Myr. We derive therefore in this section the mass function of IC\,4665
assuming ages of 50\,Myr and 100\,Myr and the concomitant interstellar
extinctions. Masses are assigned to low-mass objects by converting the $I$-band
magnitudes using the NEXTGEN and DUSTY models. At overlapping magnitudes these
two models do not assign the same mass for given magnitude. To overcome this, we
made an interpolation between the two models at the expected transition
effective temperature of 3000\,K $(I-z\simeq0.85)$. For the high-mass bins we
adopt the models without overshooting from the Padua group (Girardi et
al. 2000) and convert the $K$-band magnitude to mass. We stress that in
the derivation of the mass function presented here we do not incorporate any
correction for the cluster depth (4\% uncertainty on magnitude), undersampling
of the cluster population due to various effects (hiding, bad pixels etc.) nor
binarity (see e.g. Jeffries et al. 2004\nocite{2004MNRAS.351.1401J} for a discussion of these
various effects).

The two mass functions for the two adopted ages of IC\,4665 are presented in
Fig.\,\ref{mf} in dN/dM (Salpeter slope equals $-2.35$) and dN/dlogM (Salpeter
slope equals $-1.35$) representations. The 50\,Myr mass function in numbers is
presented in Table\,\ref{ta:imf}. Both mass functions correspond to candidate
members located within the tidal radius of $1.0^{\circ}$: high-mass stars (black
squares, histogram) from 2MASS all sky catalogues applying the same colour and
magnitude selection as in the previous subsection, and low-mass stars/BD
candidate members from our CFHT survey. For the latter we use asterisk symbols,
open circles and filled circles for candidate members found in 2s, 30s and 300s
exposures respectively. All mass bins have been statistically corrected for
background contamination. The three high-mass bins from 2MASS are multiplied by
15\%.  This factor has been derived using a 30 square degree region centred on
control field C2. The low-mass bins are corrected using factors that are valid
inside the tidal radius that have been estimated in the previous subsection for
the low-mass and BDs candidates, {\it viz.} 20\% and 40\%.
Error bars correspond to statistical uncertainties and the uncertainties in the
above correction factors. Additional correction was made for the high-mass
candidate members by applying the criteria reducing the non-member objects
that have been discussed in Sects.\,\ref{Kphot} and \ref{pm}.  We specifically
note the vertical dotted lines in all four panels of Fig.\,\ref{mf}. They
indicate the masses for which the cluster sequence intersects the background
stellar population. At these masses therefore a large relative number of
non-member contaminants is expected, for which no specific correction is
made.

The two mass functions in dN/dM representation (left column of
Fig.\,\ref{mf}) are compared to a Salpeter mass function for mass bins $\gta
1\,M_{\odot}$, and find reasonable correspondence. More crucial is the shape of
the mass function for bins $\lta 1\,M_{\odot}$. For this mass range we performed
an actual $\chi^{2}$-fit to the mass function, taking into account the 10
low-mass bins that represent the investigated mass range in this paper. For the
50\,Myr case (upper panel) we find a best linear fit for a slope of
$-0.61\pm0.13$ for a $\chi^{2}=3.6$. We have indicated this fit by a full line,
and the $1\,\sigma$ uncertainties on the fit-parameters by the two dotted lines.
For the 100\,Myr case (lower panel) the slope is less steep and the fit less
good as can be judged from Fig.\,\ref{mf}. We note that a power law relation
with a slope of $-0.6$ is approximately the slope of the mass function for
masses between 0.4 and $\sim0.1\,M_{\odot}$ found in other young open clusters
(see e.g. Bouvier et al. 2003). We thus conclude that down to
$\sim0.1\,M_{\odot}$, IC\,4665 confirms this trend of the mass function.
 
For the three mass bins corresponding to the lowest masses for which the mass
function crosses the H-burning limit the situation is more ambiguous. Either the
mass function shows a constant $-0.6$ slope well into the substellar regime
rendering two mass bins discordant, e.g. a dip at the stellar-substellar
boundary as indicated by the linear fit; or on the other hand, the mass function
flattens, rendering the last mass bin discordant. For both interpretations there
exists supporting arguments. 

In favour of the first interpretation is that such a dip in the mass function is
a recurrent feature of young open cluster. It is discussed in detail by Dobbie
et al. (2002)\nocite{2002MNRAS.335L..79D} and suggested to be due to dust
formation different from the dust formation recipe incorporated in the current
generation of cool model atmospheres. It occurs at $T_{\rm eff}\simeq 2700$\,K,
and reduces the object's luminosity at the investigated wavelengths. For the age
of IC\,4665 a dip in the mass function due to this effect is expected to be at
$\sim 0.07\,M_{\odot}$, i.e.  corresponding to the observed dip in the left
columns of Fig.\,\ref{mf}.

The argument in favour of a flattening of the mass function in dN/dM, can be
derived from the right-hand side column of Fig.\,\ref{mf}. Here we present the mass
function in the dN/dlogM representation and we have fitted the mass function
with lognormal functions. Lognormal functions are defined by an average quantity
(mass in this case) and the width ($\sigma$) of the lognormal distribution. A
lognormal prescription is found to be a good representation for the mass
function in the Pleiades (Moraux et al. 2003), Blanco\,1 and NGC\,2516. Bouvier
et al. (2005\nocite{2005MmSAI..76..265M}) reports for these three young open
clusters a similar average mass of around $0.3\,M_{\odot}$ and a similar
$\sigma$ of $\sim0.5$, producing quite a homogeneous picture of clusters in the
age range 100-150\,Myr.  The case for IC\,4665 is shown in the right-hand column of
Fig.\,\ref{mf}.  The top right panel corresponds again to a cluster age of 50\,Myr and
the mass distribution is well fitted by a lognormal function with an average
mass of $0.32\,M_{\odot}$ (full line), except for the lowest mass bin. If one
thus assumes that the complete mass function is described by a lognormal
relation, then one expects the mass function to actually flatten in a dN/dM
representation, as is suggested by the mass bins at the H-burning limit on the
left-hand side column of Fig.\,\ref{mf}.  

In Fig.\,\ref{mf}, we also compare the lognormal fit of IC\,4665 to the one of
the Pleiades (with an $\bar M=0.27\,M_{\odot}$, dashed line), and we note that
they are quite similar. In the case of the somewhat younger cluster Blanco\,1,
Bouvier et al. (2005) actually find the same average mass as we deduce here for
IC\,4665. In fact the average mass of the Pleiades is suggestively bracketed by
a higher average mass in younger clusters and lower average mass in the field
({\it viz.} $0.25\,M_{\odot}$, Chabrier 2003\nocite{2003PASP..115..763C}).

Finally we note, that despite the similar shape of the mass function in the
dN/dM representation (left column) for both adopted ages, the dN/dlogM
representations (right column) show a marked difference between the two ages. If
IC\,4665 were to be 100\,Myr old then a lognormal fit to the mass distribution
tells us that the average mass of the cluster is significantly larger than found
for clusters of such an age, {\it viz.} $0.45\,M_{\odot}$. This would indicate
that IC\,4665 for some reason might have lost a substantial fraction of its
lower mass constituent. In light of this we point out the remarkable coincidence
in space (few degrees distance, where 1 degree is about 6\,pc), age, and proper
motion of IC\,4665 with the young open cluster \object{Collinder\,359} for which
we refer to Lodieu et al. (2005)\nocite{2005subm}. We speculate that tidal
fields between the two cluster could have stripped off preferentially the
low-mass members of IC\,4665.

Be this as it may, and assuming the younger age for IC\,4665 we summarize that
the mass function of IC\,4665 be well approximated by a power law with slope
$-0.6$ between 1 and $\sim 0.1\,M_{\odot}$. The shape of the mass function of
IC\,4665 below the H-burning limit can either be explained in terms of the
``M7-8'' gap as first described by Dobbie et al. (2002) producing a dip with
respect to a linear relation, or alternatively, by assuming that mass functions
in general are lognormal functions, in which case the mass bin corresponding to
the lowest mass in Fig.\,\ref{mf} would be erroneous for a yet unknown
reason. At the moment we cannot exclude either interpretation for the shape of
the substellar mass function of IC\,4665.

\subsection{The total mass and tidal radius}
In the previous subsection, we arrived at a tidal radius for IC\,4665 of
$\sim1^{\circ}$ along various lines of analysis. The tidal radius depends
amongst others on the total cluster mass and Galactic orbital radius of the
cluster, a relation that was cast in the form of an equation for the Solar
neighbourhood by Pinfield et al. (1998). We note that, as the authors point out,
this equation is based on a simplified model. In the following we estimate the total mass of
IC\,4665 and confront this with the tidal radius.  Under the assumptions that
(1) the cluster census is complete for the high-mass stars down to $V<10$ (which
is likely to be an underestimate, see Fig.\,\ref{prevmem}), and (2) a
Salpeter-law holds between 5 and $0.7\,M_{\odot}$ and a power law with slope
$-0.6$ down to $0.01\,M_{\odot}$, we find the total cluster mass of $\sim
300\,M_{\odot}$. Here we accounted for an estimated 50\% binarity with a flat q
distribution. An upper limit of $<440\,M_{\odot}$ to the cluster mass is
furnished by assuming that all high-mass {\it candidate} member stars are actual
members.  An other and independent mass estimate can be found using 2MASS
catalogues. Selecting blue ($J-K<0.6$) high-mass stars within the tidal radius
of IC\,4665 and correct them for the background per $K$-band bin, one counts
between 4.5 and $1.15\,M_{\odot}$ a total of 56 stars. This translates to a
total cluster mass of $350\,M_{\odot}$, within the above upper limit. IC\,4665
has thus a mass which is probably about half the mass of the Pleiades (Pinfield
et al. 1998\nocite{1998MNRAS.299..955P}). Comparing now the cluster mass estimate with the tidal
radius we find that IC\,4665 does not follow the relation between cluster mass
and tidal radius of Pinfield et al. (1998), their equation 12. In fact IC\,4665 seems too small for
the lowest of the above mass estimates by $0.5^{\circ}$. The mass and tidal
radius can be reconciled in terms of the Pinfield relation if the distance of
IC\,4665 would have been underestimated by 200\,pc. It would render the distance
modulus 1 magnitude larger, which seems unlikely (see
Fig.\,\ref{prevmem}). Alternatively tidal stripping of cluster members may also
depend on the local environment of the cluster. We again point, as in the
previous paragraph to the fact that IC\,4665 is located in the Galactic field 
suspiciously close to the young open cluster \object{Collinder\,359} (Lodieu et al. 2005).  We
speculate that a small tidal radius for the total mass and a relatively large
average mass for a cluster age of 100\,Myr could possibly be both the
manifestation of ongoing tidal interactions and subsequent stripping of cluster
members of IC\,4665 and \object{Collinder\,359}. Whether this process is really
in effect certainly warrants a deeper investigation into the environment of
these two young open clusters.

\section{Summary}
\label{summary}
We have presented the photometric selection of 691 low-mass stellar and 94 BD
candidate members to the young open cluster IC\,4665. This extends the census of
cluster (candidate) members well into the BD regime, going down to a few tens of
Jupiter masses. We make an estimate of the contamination due to foreground and
background objects in the field of IC\,4665 on a statistical basis using control
fields located at the same Galactic latitude. An additional selection procedure
of contaminant objects for low-mass stellar and BD candidates is reported on the
basis of public (2MASS) and new $K$-band photometry. Public proper motion survey
data (Tycho-2, UCAC2) are also used to weed out candidate members for the
somewhat brighter objects. Despite the substantial background contamination (up
to 85\% taken over the full surveyed area) the cluster does clearly stand out on
the sky from the background population. This is the case for the brighter
objects, but also the lower mass stars and BDs clearly show an increased number
surface density towards the centre of the cluster. We study the radial
distribution of various cluster populations and find a consistent tidal radius
of $1^{\circ}$.  There could be an indication of mass segregation, but this
needs confirmation. 

We find IC\,4665 typical for a cluster in this age range with respect to (1) the
fraction of BD (down to $\sim 0.03\,M_{\odot}$) to total members, {\it viz} 10-19\%; (2) the slope of the mass function
that it is well described by a power law with a power of $-0.6$
for the low-mass objects downto $\sim0.1\,M_{\odot}$; (3) a cusp in the mass
function at about the Hydrogen burning limit, than can be explained in terms of
the ``missing'' M7-8 dwarfs, alternatively the cusp may be the product of an
actual flattening of the mass function and a spurious value for the lowest mass bin,
in concordance with the mass function being lognormal; (4) an average mass of
$0.32\,M_{\odot}$ when assuming an age of 50\,Myr; (5) the width of $\sigma=0.5$
for the lognormal function fitted to the mass function in a logarithmic
representation. However in two respects IC\,4665 may stand out. We estimate the
total cluster mass and find that it does not agree with the expected mass
derived from the relation between tidal radius and total cluster mass (Pinfield
et al. 1998). Secondly, we find that the average mass of IC\,4665 is relatively
large when assuming an age of 100\,Myr. We speculate that increased stripping of
member stars due to the interaction with the close-by young open cluster
Collinder\,359 could be the cause of these two effects. Clearly a better age
determination for IC\,4665 may resolve this issue.  The analysis presented here
adds in creating a consistent picture of the mass distribution in young open
clusters. A rigorous evaluation of the mass function of IC\,4665 using a large
body of spectroscopic data will be the subject of paper\,II in our series on the
young open cluster IC\,4665.

\begin{acknowledgements}
We would like to thank E. Bertin for providing PSFex, I. Baraffe for computing
isochrones for CFHT 12K filters, and E. Magnier for help in the derivation of
photometric ZP. The research benefitted from financial assistance from the
European Union Research Training Network `The Formation and Evolution of Young
Stellar Clusters' (RTN1-1999-00436). This research has made use of the Simbad
database, operated at the Centre de Donnees Astronomiques de Strasbourg (CDS),
and of NASA's Astrophysics Data System Bibliographic Services (ADS). This
publication makes use of data products from the Two Micron All Sky Survey, which
is a joint project of the University of Massachusetts and the Infra-red
Processing and Analysis Center/California Institute of Technology, funded by the
National Aeronautics and Space Administration and the National Science
Foundation. Finally, WJDW would like to thank the staff and students of the Osservatorio
di Arcetri in Florence for their warm hospitality throughout the years, grazie mille.
\end{acknowledgements}

\bibliographystyle{aa}

\begin{thebibliography}{47}
\expandafter\ifx\csname natexlab\endcsname\relax\def\natexlab#1{#1}\fi

\bibitem[{{Baraffe} {et~al.}(1998){Baraffe}, {Chabrier}, {Allard}, \&
  {Hauschildt}}]{1998A&A...337..403B}
{Baraffe}, I., {Chabrier}, G., {Allard}, F., \& {Hauschildt}, P.~H. 1998, \aap,
  337, 403

\bibitem[{{Baraffe} {et~al.}(2002){Baraffe}, {Chabrier}, {Allard}, \&
  {Hauschildt}}]{2002A&A...382..563B}
---. 2002, \aap, 382, 563

\bibitem[{{Barrado y Navascu{\' e}s} {et~al.}(2001){Barrado y Navascu{\' e}s},
  {Stauffer}, {Bouvier}, \& {Mart{\'{\i}}n}}]{2001ApJ...546.1006B}
{Barrado y Navascu{\' e}s}, D., {Stauffer}, J.~R., {Bouvier}, J., \&
  {Mart{\'{\i}}n}, E.~L. 2001, \apj, 546, 1006

\bibitem[{{Bertin} \& {Arnouts}(1996)}]{1996A&AS..117..393B}
{Bertin}, E. \& {Arnouts}, S. 1996, \aaps, 117, 393

\bibitem[{{Bonnell} \& {Davies}(1998)}]{1998MNRAS.295..691B}
{Bonnell}, I.~A. \& {Davies}, M.~B. 1998, \mnras, 295, 691

\bibitem[{{Bouvier} {et~al.}(2003){Bouvier}, {Moraux}, {Stauffer}, {Barrado y
  Navascu{\' e}s}, \& {Cuillandre}}]{2003IAUS..211..147B}
{Bouvier}, J., {Moraux}, E., {Stauffer}, J.~R., {Barrado y Navascu{\' e}s}, D.,
  \& {Cuillandre}, J. 2003, in IAU Symposium, 147--+

\bibitem[{{Chabrier}(2003)}]{2003PASP..115..763C}
{Chabrier}, G. 2003, \pasp, 115, 763

\bibitem[{{Chabrier} {et~al.}(2000){Chabrier}, {Baraffe}, {Allard}, \&
  {Hauschildt}}]{2000ApJ...542..464C}
{Chabrier}, G., {Baraffe}, I., {Allard}, F., \& {Hauschildt}, P. 2000, \apj,
  542, 464

\bibitem[{{Close} {et~al.}(2003){Close}, {Siegler}, {Freed}, \&
  {Biller}}]{2003ApJ...587..407C}
{Close}, L.~M., {Siegler}, N., {Freed}, M., \& {Biller}, B. 2003, \apj, 587,
  407

\bibitem[{{Cuillandre} {et~al.}(2001){Cuillandre}, {Luppino}, {Starr}, \&
  {Isani}}]{2001sf2a.conf..605C}
{Cuillandre}, J., {Luppino}, G., {Starr}, B., \& {Isani}, S. 2001, in
  SF2A-2001: Semaine de l'Astrophysique Francaise, 605--+

\bibitem[{{de Grijs} {et~al.}(2002){de Grijs}, {Gilmore}, {Johnson}, \&
  {Mackey}}]{2002MNRAS.331..245D}
{de Grijs}, R., {Gilmore}, G.~F., {Johnson}, R.~A., \& {Mackey}, A.~D. 2002,
  \mnras, 331, 245

\bibitem[{{Delgado-Donate} {et~al.}(2004){Delgado-Donate}, {Clarke}, \&
  {Bate}}]{2004MNRAS.347..759D}
{Delgado-Donate}, E.~J., {Clarke}, C.~J., \& {Bate}, M.~R. 2004, \mnras, 347,
  759

\bibitem[{{Dobbie} {et~al.}(2002){Dobbie}, {Pinfield}, {Jameson}, \&
  {Hodgkin}}]{2002MNRAS.335L..79D}
{Dobbie}, P.~D., {Pinfield}, D.~J., {Jameson}, R.~F., \& {Hodgkin}, S.~T. 2002,
  \mnras, 335, L79

\bibitem[{{Giampapa} {et~al.}(1998){Giampapa}, {Prosser}, \&
  {Fleming}}]{1998ApJ...501..624G}
{Giampapa}, M.~S., {Prosser}, C.~F., \& {Fleming}, T.~A. 1998, \apj, 501, 624

\bibitem[{{Girardi} {et~al.}(2000){Girardi}, {Bressan}, {Bertelli}, \&
  {Chiosi}}]{2000A&AS..141..371G}
{Girardi}, L., {Bressan}, A., {Bertelli}, G., \& {Chiosi}, C. 2000, \aaps, 141,
  371

\bibitem[{{H{\o}g} {et~al.}(2000){H{\o}g}, {Fabricius}, {Makarov}, {Urban},
  {Corbin}, {Wycoff}, {Bastian}, {Schwekendiek}, \&
  {Wicenec}}]{2000A&A...355L..27H}
{H{\o}g}, E., {Fabricius}, C., {Makarov}, V.~V., {et~al.} 2000, \aap, 355, L27

\bibitem[{{Hoogerwerf} {et~al.}(2001){Hoogerwerf}, {de Bruijne}, \& {de
  Zeeuw}}]{2001A&A...365...49H}
{Hoogerwerf}, R., {de Bruijne}, J.~H.~J., \& {de Zeeuw}, P.~T. 2001, \aap, 365,
  49

\bibitem[{{Hunt} {et~al.}(1998){Hunt}, {Mannucci}, {Testi}, {Migliorini},
  {Stanga}, {Baffa}, {Lisi}, \& {Vanzi}}]{1998AJ....115.2594H}
{Hunt}, L.~K., {Mannucci}, F., {Testi}, L., {et~al.} 1998, \aj, 115, 2594

\bibitem[{{Jeffries} {et~al.}(2004){Jeffries}, {Naylor}, {Devey}, \&
  {Totten}}]{2004MNRAS.351.1401J}
{Jeffries}, R.~D., {Naylor}, T., {Devey}, C.~R., \& {Totten}, E.~J. 2004,
  \mnras, 351, 1401

\bibitem[{{Jiang} {et~al.}(2004){Jiang}, {Laughlin}, \&
  {Lin}}]{2004AJ....127..455J}
{Jiang}, I., {Laughlin}, G., \& {Lin}, D.~N.~C. 2004, \aj, 127, 455

\bibitem[{{Kalirai} {et~al.}(2001){Kalirai}, {Richer}, {Fahlman}, {Cuillandre},
  {Ventura}, {D'Antona}, {Bertin}, {Marconi}, \&
  {Durrell}}]{2001AJ....122..257K}
{Kalirai}, J.~S., {Richer}, H.~B., {Fahlman}, G.~G., {et~al.} 2001, \aj, 122,
  257

\bibitem[{{King}(1962)}]{1962AJ.....67..471K}
{King}, I. 1962, \aj, 67, 471

\bibitem[{{Kroupa}(2001)}]{2001MNRAS.322..231K}
{Kroupa}, P. 2001, \mnras, 322, 231

\bibitem[{{Kroupa} \& {Bouvier}(2003)}]{2003MNRAS.346..369K}
{Kroupa}, P. \& {Bouvier}, J. 2003, \mnras, 346, 369

\bibitem[{{Lin} {et~al.}(1998){Lin}, {Laughlin}, {Bodenheimer}, \&
  {Rozyczka}}]{1998Sci...281.2025L}
{Lin}, D.~N.~C., {Laughlin}, G., {Bodenheimer}, P., \& {Rozyczka}, M. 1998,
  Science, 281, 2025

\bibitem[{{Littlefair} {et~al.}(2003){Littlefair}, {Naylor}, {Jeffries},
  {Devey}, \& {Vine}}]{2003MNRAS.345.1205L}
{Littlefair}, S.~P., {Naylor}, T., {Jeffries}, R.~D., {Devey}, C.~R., \&
  {Vine}, S. 2003, \mnras, 345, 1205

\bibitem[{{Lodieu} {et~al.}(2005){Lodieu}, {Bouvier}, {James}, {de Wit},
  {Palla}, {McCaughrean}, \& {Cuillandre}}]{2005subm}
{Lodieu}, N., {Bouvier}, J., {James}, D.~J., {et~al.} 2005, \aap

\bibitem[{{Luhman}(2004)}]{2004ApJ...614..398L}
{Luhman}, K.~L. 2004, \apj, 614, 398

\bibitem[{{Luhman} {et~al.}(2003){Luhman}, {Stauffer}, {Muench}, {Rieke},
  {Lada}, {Bouvier}, \& {Lada}}]{2003ApJ...593.1093L}
{Luhman}, K.~L., {Stauffer}, J.~R., {Muench}, A.~A., {et~al.} 2003, \apj, 593,
  1093

\bibitem[{{Magnier} \& {Cuillandre}(2004)}]{2004PASP..116..449M}
{Magnier}, E.~A. \& {Cuillandre}, J.-C. 2004, \pasp, 116, 449

\bibitem[{{Martin} \& {Montes}(1997)}]{1997A&A...318..805M}
{Martin}, E.~L. \& {Montes}, D. 1997, \aap, 318, 805

\bibitem[{{Mermilliod}(1981)}]{1981A&A....97..235M}
{Mermilliod}, J.~C. 1981, \aap, 97, 235

\bibitem[{{Mohanty} {et~al.}(2005){Mohanty}, {Jayawardhana}, \&
  {Basri}}]{2005ApJ...626..498M}
{Mohanty}, S., {Jayawardhana}, R., \& {Basri}, G. 2005, \apj, 626, 498

\bibitem[{{Moraux} {et~al.}(2005){Moraux}, {Bouvier}, \&
  {Clarke}}]{2005MmSAI..76..265M}
{Moraux}, E., {Bouvier}, J., \& {Clarke}, C. 2005, Memorie della Societa
  Astronomica Italiana, 76, 265

\bibitem[{{Moraux} {et~al.}(2003){Moraux}, {Bouvier}, {Stauffer}, \&
  {Cuillandre}}]{2003A&A...400..891M}
{Moraux}, E., {Bouvier}, J., {Stauffer}, J.~R., \& {Cuillandre}, J.-C. 2003,
  \aap, 400, 891

\bibitem[{{Natta} {et~al.}(2002){Natta}, {Testi}, {Comer{\' o}n}, {Oliva},
  {D'Antona}, {Baffa}, {Comoretto}, \& {Gennari}}]{2002A&A...393..597N}
{Natta}, A., {Testi}, L., {Comer{\' o}n}, F., {et~al.} 2002, \aap, 393, 597

\bibitem[{{Natta} {et~al.}(2004){Natta}, {Testi}, {Muzerolle}, {Randich},
  {Comer{\' o}n}, \& {Persi}}]{2004A&A...424..603N}
{Natta}, A., {Testi}, L., {Muzerolle}, J., {et~al.} 2004, \aap, 424, 603

\bibitem[{{Papaloizou} \& {Terquem}(2001)}]{2001MNRAS.325..221P}
{Papaloizou}, J.~C.~B. \& {Terquem}, C. 2001, \mnras, 325, 221

\bibitem[{{Pinfield} {et~al.}(1998){Pinfield}, {Jameson}, \&
  {Hodgkin}}]{1998MNRAS.299..955P}
{Pinfield}, D.~J., {Jameson}, R.~F., \& {Hodgkin}, S.~T. 1998, \mnras, 299, 955

\bibitem[{{Prosser}(1993)}]{1993AJ....105.1441P}
{Prosser}, C.~F. 1993, \aj, 105, 1441

\bibitem[{{Prosser} \& {Giampapa}(1994)}]{1994AJ....108..964P}
{Prosser}, C.~F. \& {Giampapa}, M.~S. 1994, \aj, 108, 964

\bibitem[{{Reipurth} \& {Clarke}(2001)}]{2001AJ....122..432R}
{Reipurth}, B. \& {Clarke}, C. 2001, \aj, 122, 432

\bibitem[{{Shu} {et~al.}(1987){Shu}, {Adams}, \&
  {Lizano}}]{1987ARA&A..25...23S}
{Shu}, F.~H., {Adams}, F.~C., \& {Lizano}, S. 1987, \araa, 25, 23

\bibitem[{{Starr} {et~al.}(2000){Starr}, {Doyon}, {Beuzit}, {Vallee}, {Calder},
  {Eriksen}, {Cuillandre}, {Grundseth}, {Barrick}, {Ward}, {Knight}, \&
  {Nadeau}}]{2000SPIE.4008..999S}
{Starr}, B.~M., {Doyon}, R., {Beuzit}, J., {et~al.} 2000, in Proc. SPIE Vol.
  4008, p. 999-1009, Optical and IR Telescope Instrumentation and Detectors,
  Masanori Iye; Alan F. Moorwood; Eds., 999--1009

\bibitem[{{Whitworth} \& {Zinnecker}(2004)}]{2004A&A...427..299W}
{Whitworth}, A.~P. \& {Zinnecker}, H. 2004, \aap, 427, 299

\bibitem[{{Zacharias} {et~al.}(2004){Zacharias}, {Urban}, {Zacharias},
  {Wycoff}, {Hall}, {Monet}, \& {Rafferty}}]{2004AJ....127.3043Z}
{Zacharias}, N., {Urban}, S.~E., {Zacharias}, M.~I., {et~al.} 2004, \aj, 127,
  3043

\bibitem[{{Zapatero Osorio} {et~al.}(1997){Zapatero Osorio}, {Martin}, \&
  {Rebolo}}]{1997A&A...323..105Z}
{Zapatero Osorio}, M.~R., {Martin}, E.~L., \& {Rebolo}, R. 1997, \aap, 323, 105

\end{thebibliography}



\end{document}